\documentclass[aps,twocolumn,prl,floatfix,superscriptaddress]{revtex4-2}

\usepackage{graphicx}
\usepackage{amsmath,amsfonts,amssymb,bm}

\usepackage{lmodern}
\usepackage{xcolor}
\usepackage[normalem]{ulem}
\usepackage{float}
\usepackage{braket}
\usepackage{dsfont}

\makeatletter
\usepackage{pdfpages} % include appendix
\AtBeginDocument{\let\LS@rot\@undefined}
\makeatother

\usepackage[colorlinks,
            linkcolor=blue,    
            citecolor=blue,  
            urlcolor=blue,
 	        bookmarks=true,        
	        bookmarksopen=true,    
	        bookmarksnumbered=true,
]{hyperref}

\newcommand{\bs}[1]{\boldsymbol{#1}}
\newcommand{\beq}{\begin{equation}}
\newcommand{\eeq}{\end{equation}}

\newcommand{\ra}{\rangle}
\newcommand{\kket}[1]{\ket{\ket{#1}}}
\newcommand{\bbra}[1]{\bra{\bra{#1}}}

\newcommand{\rra}{\ra\ra}

\newcommand{\ov}[1]{\overline{#1}}
\newcommand{\ttbar}{\Theta\bar{\Theta}}

\global\long\def\bbrakket#1#2{\left\langle \left\langle #1\right.\left\Vert #2\right\rangle \right\rangle }%
\definecolor{magenta}{rgb}{1.0, 0.0, 1.0}

\DeclareMathOperator{\tr}{tr}

\begin{document}

\title{Diffusive Operator Spreading for Random Unitary Free Fermion Circuits}

\author{Beatriz~Dias}
\affiliation{CeFEMA, Instituto Superior T\'{e}cnico, Universidade de Lisboa, Av. Rovisco Pais, 1049-001 Lisboa, Portugal}

\author{Masudul~Haque}
\affiliation{Department of Theoretical Physics, Maynooth University, Co.\ Kildare, Ireland}
\affiliation{Max Planck Institute for the Physics of Complex Systems, N\"{o}thnitzer Str. 38, 01187 Dresden, Germany}
\author{Pedro~Ribeiro}
\affiliation{CeFEMA, Instituto Superior T\'{e}cnico, Universidade de Lisboa, Av. Rovisco Pais, 1049-001 Lisboa, Portugal}
\affiliation{Beijing Computational Science Research Center, Beijing 100084, China}
\author{Paul~A.~McClarty}
\affiliation{Max Planck Institute for the Physics of Complex Systems, N\"{o}thnitzer Str. 38, 01187 Dresden, Germany}

\begin{abstract}
We study a model of free fermions on a chain with dynamics generated by random unitary gates acting on nearest neighbor bonds and present an exact calculation of time-ordered and out-of-time-ordered correlators. We consider three distinct cases: the random circuit with spatio-temporal disorder (i) with and (ii) without particle number conservation and (iii) the particle non-conserving case with purely temporal disorder. 
In all three cases, temporal disorder causes diffusive operator spreading and $\sim\sqrt{t}$ entanglement growth. This is in sharp contrast to Anderson localization for the case of static disorder and with the ballistic behavior observed in both the clean case of Hamiltonian evolution and in fully random unitary quantum circuits.
\end{abstract}

\maketitle

{\it Introduction} $-$ Understanding the nature of quantum dynamics in many-body interacting systems far from equilibrium is one of the central issues in physics. Removed from the sanctuary of linear response theory, one must often resort to numerical simulations but the exponential scaling of Hilbert space dimension typically limits these to small system sizes. Even so, huge progress has been made over recent years driven partly by experiments on quantum simulators, numerical developments and occasional exact calculations.

In the spirit of looking for universal features in many-body dynamics, random unitary circuits have been intensively studied in the last few years (see, for example, \cite{PhysRevX.8.021014,Nahum2017,PhysRevX.8.021013,PhysRevX.8.031057,PhysRevX.8.031058,PhysRevB.100.125139,PhysRevX.8.041019}). These models describe space and time processes through the application of local random unitary gates to some underlying degrees of freedom. Since there is no Hamiltonian dynamics, even energy conservation is sacrificed in order to uncover generic features of local dynamics. Since traditional correlators are randomized at each discrete time step, the strength of these models is in capturing the spread of quantum information including the initial linear growth of bipartite entanglement from an initial product state to a fully random state with Page entanglement. One observable that is particularly suited to such models is the degree to which spatially separated local operators commute after time evolution:
\beq
\mathcal{C}(r,t)\equiv \frac{1}{2}{\rm Tr}\left[ \rho \left[ \mathcal{O}_0(t) , \mathcal{O}_r \right]^\dagger  \left[ \mathcal{O}_0(t) , \mathcal{O}_r \right] \right]
\label{eq:otoc}
\eeq
where $\mathcal{O}_r$ is an operator localized at position $r$ \cite{1969JETP...28.1200L,swingle2018unscrambling,2017JHEP...10..138H,maldacena2016bound,kitaev2015,hosur2016chaos,Fan_2017,PhysRevB.98.144304,PhysRevA.94.040302,garttner2017measuring,PhysRevX.7.031011,mi2021information}. Expanding out the correlator gives both time ordered correlators (TOC) and out-of-time-ordered correlators (OTOC). The quantity $\mathcal{C}(r,t)$ is known to exhibit ballistic spreading and KPZ growth at the light cone interface in even the simplest variant of random circuit models \cite{PhysRevX.8.021014,PhysRevX.8.021013} and the dynamics can be mapped to a biased random walk. In the presence of a conserved charge, the picture is modified owing to the diffusion of the conserved charges \cite{PhysRevX.8.031057,PhysRevX.8.031058}. Thus, one finds there is a ballistic front that itself spreads  diffusively. As with everything else in random matrix theory, the importance attached to the OTOC in random unitary circuits is that it is conjectured to provide a tractable instance of universal physics in this case tied to thermalization and the scrambling of quantum information in general non-integrable many-body interacting systems \cite{PhysRevB.101.024202,PhysRevB.99.174203,PhysRevB.98.134305,PhysRevB.99.054205,2020arXiv200510257C}.

\begin{figure}[H]%[tp]
  \centering
  \includegraphics[width=\columnwidth]{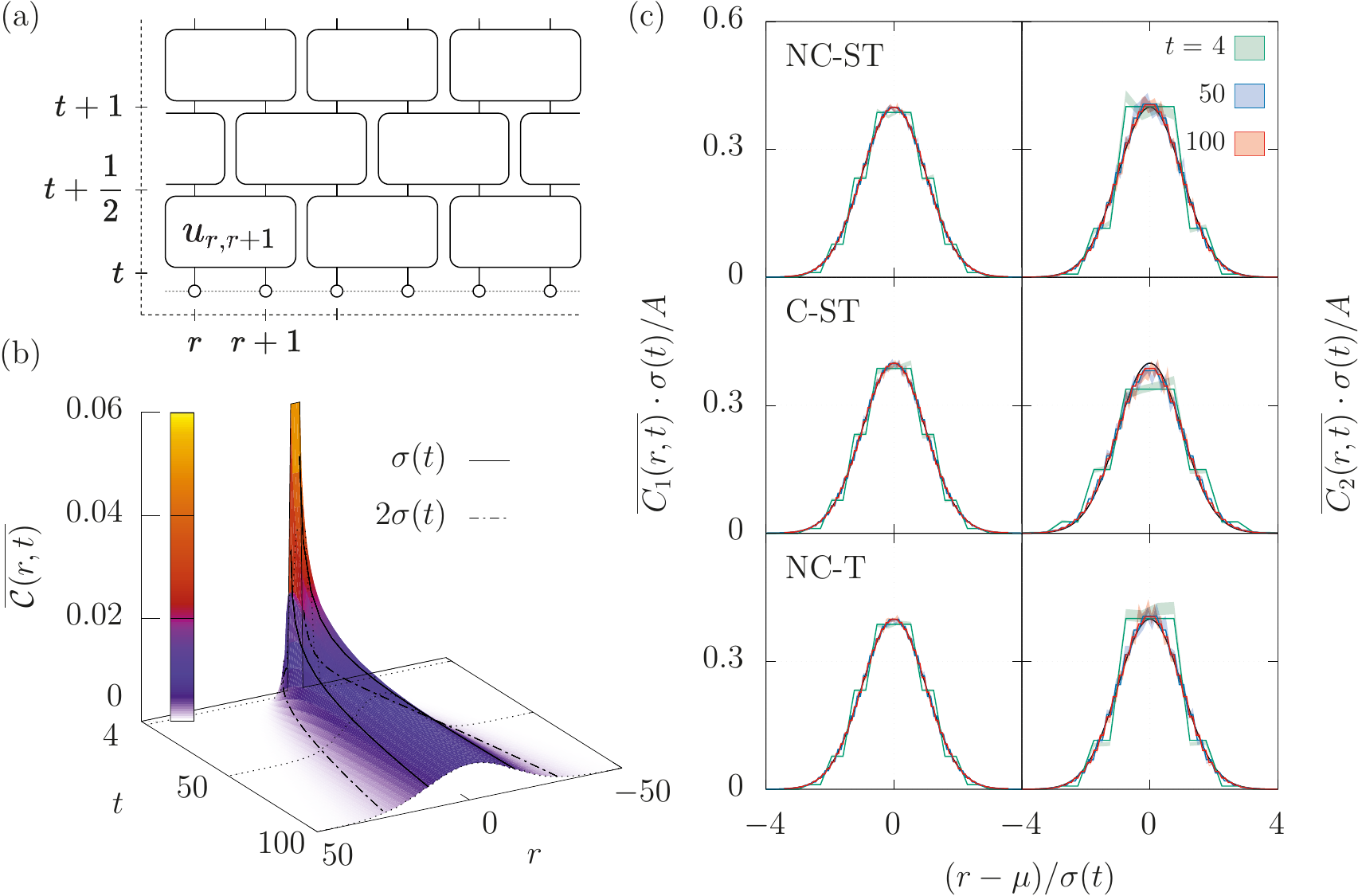}
  \caption{\label{fig:diffusion} 
(a) Circuit scheme in the $r-t$ plane: the random two-site unitaries $u_{r,r+1}$ are applied at neighboring sites in a brick wall pattern. 
(b) Exact $\overline{\mathcal{C}(r,t)} = 1/2^3[\overline{C_1(r,t)}-\overline{C_2(r,t)}]$ as an height and color map for a system with $100$ sites in the case of NC-ST (and NC-T) evolution. The black curves envelop the $\sigma(t)=\sqrt{2t}$ and $2\sigma(t)=\sqrt{8t}$ regions for $\overline{C_1}$, which is the leading order term in $\overline{\mathcal{C}(r,t)}$.
(c) Rescaled TOC (left) and OTOC (right), i.e. $\ov{C_1(r,t)}\, \sigma(t)/A$ and $\ov{C_2(r,t)}\, \sigma(t)/A(t)$ (with $A$ the normalization), as a function of $(r-\mu)/\sigma(t)$ for three fixed time steps $t\in\{4,50,100\}$ and for NC-ST (top), CS-T (middle) and NC-T (bottom). Full lines correspond to exact calculations (shown in (a) for NC-ST) and the color filled region to simulations which include an average of $4000$ disorder realizations, starting from a random state with particle number fixed to $L/2$. Both collapse to the continuum limit solution, i.e. the Gaussian $g(r,t) = 1/(\sqrt{2\pi}) \exp[- (r-\mu)^2/(2\sigma^2(t))]$. For $\ov{C_1(r,t)}$, $A=2$ and $\sigma(t)=\sqrt{2t}$ for all cases; for $\ov{C_2(r,t)}$, $\sigma(t)=\sqrt{t}$ for all cases, $A(t)=1/(2\sqrt{2\pi t})$ for NC-ST and NC-T and $A(t)=2/(\sqrt{2\pi t})$ for C-ST. The deviation $\mu=1/2$ centers $\overline{C_1}$ and $\overline{C_2}$ around $r=0$.
}
\end{figure}

In this paper we consider driven free fermion dynamics within the context of random unitary circuits. Free fermion circuits first appeared as classically simulatable matchgate circuits \cite{valiant2002quantum}, which were later shown to correspond to a model of free fermions in 1D \cite{Terhal_2002}.
Unlike their many-body interacting analogues, there is evidence that free fermion models with spatio-temporal noise exhibit diffusive dynamical features that were observed in the growth of the von Neumann entropy  \cite{Nahum2017, Roosz2016}, in large-deviation statistics  \cite{Bernard2019} and in the magnetization dynamics of the transverse field Ising model that maps to a free fermion problem. 
The presence of diffusive dynamics in such cases is a nontrivial result that sits in contrast to the Anderson localization expected for static spatial disorder or to the ballistic behavior expected in the clean case. 
The extent to which interactions change this picture is not yet fully understood. Nonetheless, in generic (i.e. interacting) quantum random circuits both OTOCs and the entanglement dynamics spread ballistically.   

In the following, we study the effect of spatio-temporal noise on the dynamics of free fermions in a setting where an {\it exact} calculation of the $\mathcal{C}(r,t)$ for a random circuit is possible. 
In this paper, we establish through an exact calculation the diffusive behavior from the underlying dynamics of $\mathcal{C}(r,t)$ for quadratic fermions.
We consider three distinct instances of free fermion evolution: a particle conserving spatio-temporal random circuit (C-ST), its generalization to a non particle conserving process (NC-ST), and a spatially homogeneous case where randomness appears only in the time direction (NC-T). In the Supplementary Section, we also present numerical results for a non particle conserving circuit with quenched spatial disorder (NC-S) that Anderson localizes.

The random circuit acts on a chain of $L$ sites $\left(r=0,...,L-1\right)$ and periodic boundary conditions as a discrete time protocol. A time step corresponds to the action of the unitary operation $\mathcal{U}=\mathcal{U}_{\text{even}}\mathcal{U}_{\text{odd}}$, obtained by the successive application of the half-steps $\mathcal{U}_{\text{even}}=\prod_{R=0}^{L/2-1} \mathcal{U}_{2R,2R+1}$ and $\mathcal{U}_{\text{odd}}=\prod_{R=0}^{L/2-1} \mathcal{U}_{2R-1,2R}$.
Here, $\mathcal{U}_{r,r+1}$ is a random unitary acting nontrivially on the local 4-dimensional Hilbert space of sites $r$ and $r+1$.  The evolution follows the brick wall pattern shown of Fig.~\ref{fig:diffusion}(a). 
For the free fermion evolution, instead of sampling $\mathcal{U}_{r,r+1}$ over U$(4)$ \cite{Nahum2017}, we consider a sub-group that leaves invariant the algebra of single-particle fermionic creation and annihilation operators ($a_{r}^{\dagger}$ and $a_{r}$) \cite{valiant2002quantum,Terhal_2002}, i.e. $\mathcal{U}^{\dagger}_{r,r+1}{\Psi}_{r}\mathcal{U}_{r,r+1}=u_{r,r+1}{\Psi}_{r}$, where $\Psi_{r}\equiv(a_{r},a_{r+1},a_{r}^{\dagger},a_{r+1}^{\dagger})^{T}$ and $u_{r,r+1}$ is a $4\times 4$ unitary matrix respecting particle-hole symmetry,
$\tau_{1}u_{r,r+1}^{T}\tau_{1}=u_{r,r+1}^{\dagger}$, with $\tau_{1}$ a Pauli matrix acting on Nambu space. When $\det u_{r,r+1}=1$, such unitary operator can be obtained by continuous time evolution under a quadratic (free) fermionic Hamiltonian during a period of time $\Delta t$, i.e. $\mathcal{U}_{r,r+1}=\exp(-\frac{i}{2}\Psi_{r}^{\dagger}H_{r}\Psi_{r}\Delta t)$, in which case $u_{r,r+1}=\exp(-iH_{r}\Delta t)$. Here we take $u_{r,r+1}$ to be Haar-distributed \cite{SM}. 
Note that for NC-T at each time step the same is applied to all pairs of sites, i.e. $u_{r,r+1} = u$, and for C-ST $u_{r,r+1} = v_{r,r+1} \oplus v^*_{r,r+1} $, with $v_{r,r+1}$ a $2\times2$ Haar-distributed unitary matrix. 

\begin{figure}[tp]
  \centering
  \includegraphics[width=0.95\columnwidth,angle=0]{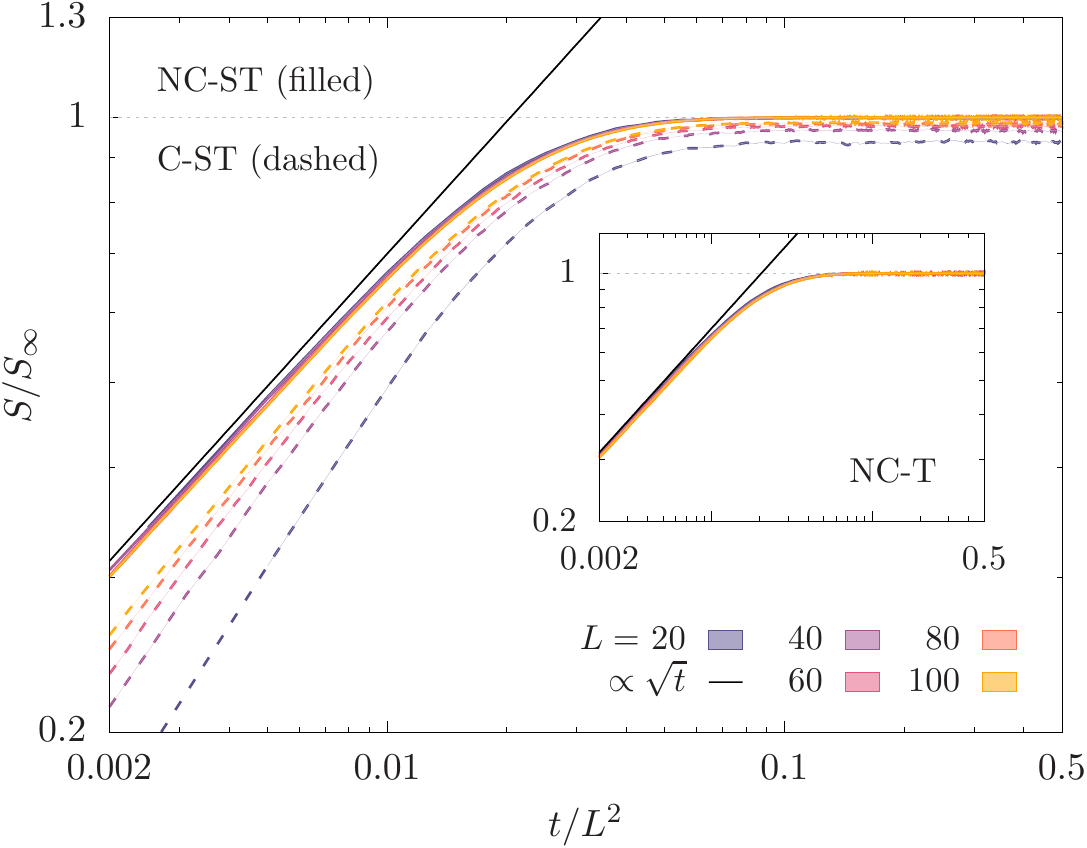}
  \caption{\label{fig:SvN}
Growth of the entanglement entropy divided by the saturation value $S/S_\infty$ as a function of $t/L^2$ for random free fermions $-$ NC-ST (filled), C-ST (dashed) and NC-T (inset) $-$ with $L=100$ and a subsystem of $50$ sites showing $S(t)\propto\sqrt{t}$ at the earliest times. 
The data results from averaging over 1000 realizations up to $t=5000$,  starting from a random product state with particle number fixed to $L/2$. The curves for NC-ST and NC-T overlap.}
\end{figure}

{\it Dynamics of Entanglement} $-$ 
For completeness, we start by analysing the dynamics of the von Neumann entropy, $S = - \tr (\tilde{\rho} \ln \tilde{\rho})$, with $\tilde{\rho}$ the reduced density matrix of a subsystem of size $L/2$, starting from an initial product state with a well defined particle number. 
Fig.~\ref{fig:SvN} shows that for the three processes the entanglement grows as $\sim D_S \sqrt{t}$ (with $D_S$ a time independent constant) for small times saturating at times $t_{\rm sat}\sim L^2$. Asymptotically diffusive growth of the {\it R\'{e}nyi entropy} $S_2$ has been  discussed in the literature, coinciding with linear growth of the von Neumann entropy at least in low dimensional local Hilbert spaces in models with conserved quantities \cite{PhysRevLett.122.250602,znidaric_2020}. In these cases, operator growth has a ballistically propagating front with diffusive growth in the rest frame of the front. The random circuit free fermion model is therefore qualitatively different to these cases.
For $t \gg t_{\rm sat}$, the saturation value $S_\infty =  s_0  L  + s_1 + O(1/L)$, coincides with the mean entanglement entropy of a random Gaussian state \cite{Magan2016, Liu2018}, with $s_0 \simeq 0.193$ for all the considered free fermion processes and $s_1 \simeq 0.085$ for the NC-ST case, well below the Page value $(s_0 = \ln 2/2 \simeq 0.346,  s_1 = - 1/2)$ obtained by averaging over the full Hilbert space. 
These results show that the rate of increase of the entanglement is compatible with diffusion of quantum information. 
In addition, while the saturation entanglement has volume law scaling, free fermion dynamics cannot explore the full Hilbert space. 
We now turn to the signatures of free fermion dynamics in the spatial correlations.

{\it Operator Spreading} $-$
The time-ordered density-density correlator becomes trivial when averaged over temporal disorder (see Supplementary Section \cite{SM}) but $\mathcal{C}(r,t)$, introduced in Eq.~\eqref{eq:otoc}, remains non-trivial upon averaging.  
We consider an average over separable initial states, which is equivalent to taking $\rho \propto 1 $ in Eq.~\eqref{eq:otoc}, i.e. the infinite temperature ensemble. 
In the following, we shall consider quadratic observables $\mathcal{O}_r = 1/2 \Psi^\dagger O_r \Psi$ where $\Psi \equiv (a_1, \ldots, a_L, a_{1}^\dagger, \ldots, a_{L}^\dagger)^T$ and $O_r$ is a local single-particle operator. 
The computation of $\mathcal{C}(r,t)$ for free fermions - in common with other correlators - can be brought into a form where the trace need only be performed over $2L\times 2L$ matrices rather than over the entire Hilbert space. 
One can show that the many-body correlator can be written in terms of single-body quantities as $\mathcal{C}(r,t)=1/2^3[C_1(r,t)-C_2(r,t)]$ where 
\begin{align}
C_1(r,t) & =  {\rm tr}\left[ O^2_0(t) O^2_r \right], \label{eq:sp_c1} \\
C_2(r,t) & =  {\rm tr}\left[ O_0 (t)  O_r  O_0(t)  O_r  \right] ,
\label{eq:sp_otoc}
\end{align}
and $O(t) = U O(t-1) U^\dagger$. The general relation between the single particle and many-body TOC and OTOC is given in the Supplementary Material \cite{SM}.

{\it TOC and OTOC Numerics} $-$ The TOC ($C_1$) and OTOC ($C_2$) may now be simulated efficiently by generating local pseudo-random gates $u_{r,r+1}$ and averaging the result over many realizations of the time evolution. 
As before, we average over initial states that are product states of definite particle number. 
Results for $L=100$ and $4000$ disorder realizations are shown in Fig.~\ref{fig:diffusion}(c) for a symmetrized particle number operator 
$\mathcal{O}_r = 1/2(a_r^\dagger a_r - a_r a_r^\dagger) =  a_r^\dagger a_r - 1/2$. 
In contrast to the ballistic spreading seen in Haar-random circuits, $\overline{\mathcal{C}(r,t)}$ for random free fermions diffuses. This can be seen most clearly in Fig.~\ref{fig:diffusion}(c) where $\overline{C_1}\, \sigma(t)/A$ and $\overline{C_2}\,\sigma(t)/A(t)$ are shown to collapse to a Gaussian with standard deviation growing respectively as $\sqrt{2t}$ and $\sqrt{t}$ at multiple fixed times, with the horizontal axis rescaled to $(r-\mu)/\sigma$.
Examining the two terms $\overline{C_1}$ and $\overline{C_2}$ separately reveals that both the ordinary dynamical two-time correlator $\overline{C_1}$ and the true out-of-time-ordered component $\overline{C_2}$ are diffusive. The magnitude of $\overline{C_1}$ is, however, much larger than that of $\overline{C_2}$ at fixed time. These results are compatible with the observed $\sqrt{t}$ early time entanglement growth.

{\it Exact Calculation of $\overline{C_1(r,t)}$} $-$ We now look for the origin of the diffusive behavior of $\overline{\mathcal{C}(r,t)}$ examining $\overline{C_1(r,t)}$ and $\overline{C_2(r,t)}$ in turn and proceeding analytically by computing the exact averages over the random unitaries for the NC-ST case referring to the Supplementary Section \cite{SM} for further details. In the single particle picture, we may denote states of the system by $\ket{\alpha} \equiv \ket{r_\alpha, s_\alpha} \equiv \ket{2R_{r_\alpha}+b_\alpha,s_\alpha}$, where $r_\alpha=0,\ldots,L-1$ runs over the sites, $s_\alpha\in\{p,h\}$ labels the particle-hole index; $R_{r_\alpha}=0,\ldots,L/2-1$ labels the pair of sites $(r_\alpha,r_\alpha+1)$ and $b_{r_{\alpha}}\in\left\{ 0,1\right\} $ such that $r_{\alpha}=2R_{r_{\alpha}}+b_{r_{\alpha}}$. Note that the sites are acted upon by modulo $L$, due to periodic boundary conditions.

Starting with $\overline{C_1(r,t)}$, we use a notation in which operators are rendered as state vectors
$C_{1}(r,t) =\tr\left[O_{0}^{2}(t)O_{r}^{2}\right] = \bbra{O_{r}^{2}} O_{0}^{2}(t) \rra$,
where $\kket{O_{r}^{2}} \equiv \kket{O_{r}^{2}(0)}$, $\kket{O_{r}^{2}(t)} = \sum_{\alpha\beta}\kket{\alpha\beta}\left\langle \alpha\left|O_{r}^{2}(t)\right|\beta\right\rangle$ and $\kket{\alpha\beta} = \ket{\alpha}\otimes\bra{\beta}^T$.
Since $O_{r}$ is the symmetrized number operator, we get 
\begin{align}
\kket{O_{r}^{2}} =\sum_{s} \kket{r,s;r,s}.
\end{align}
For a single realization,  $O_{0}^{2}(t+1)=U^{\dagger}O_{0}^{2}(t)U$, which translates to $\kket{O_{0}^{2}(t+1)}=U^{\dagger}\otimes U^{T}\kket{O_{0}^{2}(t)}$. Applying one layer of our circuit, i.e. $U=U_{\rm{even}} U_{\rm{odd}}$, and averaging over multiple realizations of the random circuit as summarized in the Supplementary Section one finds
\begin{align}
&	\kket{\overline{O_{0}^{2}(t+1)}} = \hat{W}_1 \kket{\overline{O_{0}^{2}(t)}} \nonumber \\
& \hat{W}_1 \equiv \left(\sum_{R=0}^{L/2-1} \kket{\phi_{2R}}\bbra{\phi_{2R}}\right)  \left(\sum_{R=0}^{L/2-1} \kket{\phi_{2R-1}}\bbra{\phi_{2R-1}}\right)
, \label{eq:O02evolutionaveraged}
\end{align}
having introduced 
\begin{align}
	\kket{\phi_{r}} \equiv \frac{1}{2} \sum_{x\in\left\{ r,r+1\right\}} \sum_{s} \kket{x,s;x,s}.
\end{align}
With the initial condition $\kket{\ov{O_0^2(0)}}=\sum_s \kket{0,s;0,s}$, the complete time evolution of $\overline{C_1}$ is determined. The subspace spanned by ${ \kket{\phi_{r}} }$ is closed under the evolution thus respecting the particle-hole symmetry. These vectorized operators have the property $\bbrakket{\phi_r}{\phi_{r'}} = \delta_{r',r} + \frac{1}{2}\delta_{r',r\pm1}$ that leads to the simplification of the recursion relation, Eq.~\eqref{eq:O02evolutionaveraged}, to 
\begin{align}
	& \bbrakket{\phi_r}{\ov{O_0^2(t+1)}} \nonumber \\ & = \frac{1}{4} \Big( \bbra{\phi_{r-2}} + 2\bbra{\phi_{r}} + \bbra{\phi_{r+2}} \Big) \kket{\ov{O_0^2(t)}}. \label{eq:c1recursion}
\end{align}
From this and the initial condition we obtain
\begin{align}
& \ov{C_{1}\left(2R,t+1\right)} = \ov{C_{1}\left(2R+1,t+1\right)} \nonumber \\
& \frac{1}{4}\big( \ov{C_{1}\left(2R-2,t\right)} + 2\ov{C_{1}\left(2R,t\right)} + \ov{C_{1}\left(2R+2,t\right)} \Big) \label{eq:C1evolution}
\end{align}
for $t\geq 1$ and $\frac{1}{2} (\delta_{R-1,0} + \delta_{R,0})$ for $t=0$. Taking the continuum limit, $\lim_{t,L\rightarrow\infty} \overline{C_1(r,t)} = \lim_{a\rightarrow 0} a \overline{C'_1(x=ra, \tau = t a^2)}$,
leads to the 1D diffusion equation $\partial_\tau \overline{C'_1(x,\tau)} =  \partial_{x}^2 \overline{C'_1(x,\tau)}$ with diffusion constant $D=1$ which approximates the exact discrete time evolution very well even for relatively small times.

{\it Exact Calculation of $\overline{C_2(r,t)}$} $-$ The calculation of $\overline{C_2(r,t)}$ proceeds analogously to that of $\overline{C_1(r,t)}$ but is more involved not least because the disorder average is carried out over a product of four unitaries rather than two in the case of $\overline{C_1}$. As before, we write the correlator in a vectorized notation 
\begin{align}
C_{2}(r,t) & =\tr\left[O_{0}(t)O_{r}O_{0}(t)O_{r}\right]= \bbra{ Q_{r} } S \kket{Q_{0}(t)}, \label{eq:C2eq1}
\end{align}
with $Q_r \equiv O_r \otimes O_r$ fixed by the choice of observable to be
\begin{align}
\kket{Q_{r}} & =\sum_{s}\kket{rs,rs,rs,rs} - \kket{rs,r\bar{s},rs,r\bar{s}}. \label{eq:initial2}
\end{align}
with $s=p,h$ and the corresponding $\bar{s}=h,p$
and
\begin{align}
\kket{Q_{r}(t)} & =\sum_{\alpha\beta\mu\nu}\kket{\alpha\beta\mu\nu}\left\langle \alpha \beta \left|Q_{r}(t)\right| \mu \nu \right\rangle, \\
S & = \sum_{\alpha\beta\mu\nu} \kket{\alpha\beta\mu\nu} \bbra{\alpha\beta\nu\mu},
\end{align}
where $\kket{\alpha\beta\mu\nu} = \ket{\alpha\beta} \otimes \bra{\mu\nu}^T$. The $\kket{Q_0(t)}$ state evolves as
\begin{align}
\kket{Q_{0}(t+1)} &= U^{\dagger}\otimes U^{\dagger}\otimes U^{T} \otimes U^{T} \kket{Q_{0}(t)}
\end{align}
and an average is taken over different realizations eventually leading to the recursion relation
\begin{align}
& \bbrakket{\ttbar_{r,r'}}{\ov{Q_0(t+1)}} = \hat{W}_2 \kket{\ov{Q_0(t)}}, \label{eq:Q0Evol} \\
& \hat{W}_2 \equiv {\rm tr}\left[  \bs{M}_{r,r'} \bs{\Upsilon}_{r,r'}\right], \nonumber \\
& \bs{\Upsilon}_{r,r'} \equiv \begin{pmatrix}
\bbra{\ttbar_{r-2,r'-2}} & \bbra{\ttbar_{r-2,r'}} & \bbra{\ttbar_{r-2,r'+2}} \\
\bbra{\ttbar_{r,r'-2}} & \bbra{\ttbar_{r,r'}} & \bbra{\ttbar_{r,r'+2}} \\
\bbra{\ttbar_{r+2,r'-2}} & \bbra{\ttbar_{r+2,r'}} & \bbra{\ttbar_{r+2,r'+2}} 
\end{pmatrix}, \nonumber  
\end{align}
analogous to Eq.~\eqref{eq:c1recursion}. In the above expression, $\bs{M}_{r,r'}$ are $3\times 3$ matrices of constant coefficients given explicitly in the Supplementary Section. There are four distinct $\bs{M}_{r,r'}$ for $r'=r, r\pm 2, r\pm 4$ and one for all remaining $r'$. The 
\begin{align}
\kket{\ttbar_{r,r'}} = \kket{\Theta_{r,r'}} - \kket{\bar{\Theta}_{r,r'}}, 
\end{align}
and 
\begin{align}
\kket{\Theta_{r,r'}} &= \frac{1}{\sqrt{g_{r,r'}}} \sum_{\substack{r_\alpha \in \{r,r+1\} \\ r_\beta \in \{r',r'+1\}}} \sum_{s_\alpha, s_\beta} \kket{\alpha \beta \beta \alpha}, \label{eq:theta} \\
\kket{\bar{\Theta}_{r,r'}} &= \frac{1}{\sqrt{g_{r,r'}}} \sum_{\substack{r_\alpha \in \{r,r+1\} \\ r_\beta \in \{r',r'+1\}}} \sum_{s_\alpha, s_\beta} \kket{\alpha \bar{\alpha} \beta \bar{\beta}}, \label{eq:thetabar} 
\end{align}
with $g_{r,r'}=N(N-\delta_{r,r'})$. This completely determines the evolution of the single particle OTOC. With the initial condition, Eq.~\eqref{eq:initial2}, one finds
\begin{align}
	\ov{C_2(r,t=0)} &  = 2 \delta_{r,0}, \\
	 \ov{C_2(r,t=1)} & = \frac{1}{18}\left( \delta_{r,-1}+\delta_{r,0} \right),
\end{align}
and, for subsequent times, we define
\begin{align}
	K_{r,r'}(t) \equiv \frac{1}{2\sqrt{3}} \bbrakket{\ttbar_{r,r'}}{\ov{Q_0(t)}}, \label{eq:Kdef}
\end{align}
and use Eq.~\eqref{eq:Q0Evol} to get $\ov{C_2(2R,t+1)} = \ov{C_2(2R+1,t+1)}$ $= {\tr}\left[ \bs{M}_{2R,2R'} \bs{K}_{2R,2R'}(t)\right]$ with
\begin{align}
	%&\ov{C_2(2R,t+1)} = \ov{C_2(2R+1,t+1)} \nonumber \\
	%& = {\tr}\left[ \bs{M}_{2R,2R'} \bs{K}_{2R,2R'}(t) \right] \hspace{0.2cm} {\rm with} \nonumber \\
	& \bs{K}_{2R,2R'}(t) \equiv	\begin{pmatrix}
			K_{2R-2,2R'-2} & K_{2R-2,2R'} & K_{2R-2,2R'+2} \\
			K_{2R,2R'-2} & K_{2R,2R'} & K_{2R,2R'+2} \\
			K_{2R+2,2R'-2} & K_{2R+2,2R'} & K_{2R+2,2R'+2}
		\end{pmatrix}.
	\label{eq:C2evol}
\end{align}
The evolution thus described exactly reproduces the numerical results described above. We may now take the continuum limit from Eq.~\eqref{eq:C2evol}. This highlights one important distinction between the $\overline{C_1}$ and $\overline{C_2}$ cases: the evolution of $\overline{C_2}$ depends on the matrix $\bs{K}$ whose elements are indexed by a pair of spatial coordinates. 
In the continuum limit
$\overline{C_2(r,t)} = \lim_{a\rightarrow 0} \overline{C'_2(x=ra,\tau=ta^2)} \simeq f a^2 K'_{x,x'}(\tau)$, where $K'_{x,x'}(\tau)$ can be shown to obey the 2D diffusion equation $\partial_\tau K'_{x,x'}(\tau)=\left(\partial_x^2 + \partial_{x'}^2 \right)K'_{x,x'}(\tau)$ with initial condition $K'_{x,x'}(0)=2\delta(x)\delta(x')$ and $f=1/2$ for NC-ST \cite{SM}.
Note that 
$\lim_{t\rightarrow\infty} \overline{C_2(r,t)}/\overline{C_1(r,t)} = 0$, i.e. for large times the $\overline{C_1(r,t)}$ dictates the leading behavior of the OTOC. %$\lim_{t\rightarrow\infty} \overline{\mathcal{C}(r,t)} = \overline{C_1(r,t)}/2^3$.

{\it Extensions to C-ST, NC-T and NC-S} $-$ We have shown that both $\overline{C_1}$ and $\overline{C_2}$ spread diffusively for free fermions in 1D in the presence of spatio-temporal noise (NC-ST). We now consider exact calculations for two further cases: C-ST where the fermion particle number is conserved and each gate in the quantum circuit is chosen randomly, and NC-T where the unitary evolution is spatially homogeneous but where there is temporal noise $-$ a single gate is chosen randomly at each time step and applied to all pairs of sites. Fig.~\ref{fig:SvN} shows that the von Neumann entropy grows like $\sqrt{t}$ for all three cases: NC-ST, C-ST, NC-T. The Supplementary Section lays out in detail exact calculations of 
$\overline{C_1}$ and $\overline{C_2}$, analogous to the calculation summarized above for NC-ST, but with the continuum limit of $\overline{C_2}$ for C-ST having a different normalization due to $f=2$. The result is that there is diffusive spreading in all three cases with diffusion constants coinciding with those found for NC-ST. 
In contrast, numerical results obtained for the temporal homogeneous case, NC-S, presented in \cite{SM}, where even and odd layers of random gates are fixed and applied repeatedly in time, show that $\overline{C_1}$ and $\overline{C_2}$ remain Anderson localized, decaying exponentially around $r=0$ \cite{lagendijk2009fifty}.

{\it Conclusions} $-$ There is evidence that Hamiltonian models of quadratic fermions in one dimension exhibit diffusive spreading of correlations when subjected to noise \cite{Bernard2019,bernard2020solution}. Here we have found an analytically solvable instance of this physics in the OTOC of a random circuit model with both number conserving and non-conserving quadratic fermionic terms. In the long time limit, the states that result from this dynamics are extended with volume law entanglement but depart significantly from random matrix eigenstates.  In contrast, previously studied random unitary circuit models exhibit ballistic spreading of correlations with entanglement approaching the Page value asymptotically. All at once, this strongly suggests that, in Hamiltonian models of free fermions, Anderson localization is destroyed by the coherent noise we have considered and that ballistic propagation, expected for spatially homogeneous systems, becomes diffusive in the presence of noise. Both implications lay bare unusual features of quadratic fermion systems.

\begin{acknowledgments}
BD acknowledges support by FCT through Grant No.\ UIDB/04540/2020. BD and PR acknowledge support by FCT through Grant No.\  UID/CTM/04540/2019. 
\end{acknowledgments}

\bibliography{otoc2}
\bibliographystyle{apsrev4-1}

%%%%%%
% supplement
%%%%%%

\bigskip

\newpage
\onecolumngrid

\newpage


\section*{Supplementary material}

# *Supplementary Material :* Diffusive Operator Spreading for Random Unitary Free Fermion Circuits

Beatriz Dias,[1] Masudul Haque,[2,3] Pedro Ribeiro,[1,4] and Paul A. McClarty[3]

[1]*CeFEMA, Instituto Superior Técnico, Universidade de Lisboa, Av. Rovisco Pais, 1049-001 Lisboa, Portugal*
[2]*Department of Theoretical Physics, Maynooth University, Co. Kildare, Ireland*
[3]*Max Planck Institute for the Physics of Complex Systems, Nöthnitzer Str. 38, 01187 Dresden, Germany*
[4]*Beijing Computational Science Research Center, Beijing 100084, China*

The following supplementary section contains complete details of the computation of the TOC and OTOC for the three types of random free fermion circuit NC-ST, C-ST and NC-T together with further numerical results supporting those given in the main text.

**CONTENTS**

## I. OUTLINE

An analytical expression for the OTOC averaged over disorder realizations for evolution under generic unitary circuits was obtained by Nahum *et al.* in Ref. [1]. We present analogous analytical results for three instances of free fermion (FF) evolution: a particle conserving spatio-temporal random circuit (C-ST), its generalization to a particle non-conserving process (NC-ST) and a spatial homogeneous case where randomness appears only along the time direction (NC-T). In the end, we obtain a diffusive picture for the spreading of both a time-ordered (TOC) and out-of-time-order correlators in the $r - t$ plane, unlike the usual ballistic picture shown in Ref. [1].

The calculation detailed in this supplementary section runs over many pages so, here, we give a summary of the main steps to guide the reader through the remainder of the section. The analytical calculation of the OTOC given here is made possible firstly because the evolution of quadratic fermions can be carried out within the $2L$ dimensional space of single particle states as opposed to the full $2^L$ many-particle Hilbert space as outlined in Section II. Random quadratic fermion unitary gates are applied to neighboring pairs of sites with a single time step being composed of layers of gates applied first to odd and then to even bonds (Section III). We study a correlator of commutators $\mathcal{C}(r,t)$ (Eq. (1) in the main text) acting on the many body states that is written in terms of a time-ordered correlation function (TOC), $\mathcal{C}_1(r,t)$, and an out-of-time order correlator (OTOC), $\mathcal{C}_2(r,t)$. For our problem these can be analysed in the local single particle basis. It turns out (Section IV) that the many-body correlator (Eq. S7) can be written in terms of single particle correlators $C_1(r,t)$ and $C_2(r,t)$ which are similar respectively to $\mathcal{C}_1(r,t)$ and $\mathcal{C}_2(r,t)$. However, despite the similarity, the correspondence between $C_1$ and $\mathcal{C}_1$ (and $C_2$ and $\mathcal{C}_2$) is not one-to-one, with the single particle TOC and OTOC being a mixture of the many-body TOC and OTOC. To compute $\mathcal{C}$, we take the local operator at position $r$ to be the symmetrized fermionic number operator.

Observables for the random circuit must be disorder averaged. A summary of results on random matrix averages (needed for the average of free fermion gates) is given in Section V and further explained throughout the text. One can imagine building the random circuit for a given disorder realization, computing the desired correlator and then averaging over realizations. This computation can be done instead by studying first the action of a single layer (acting on even or odd pairs of sites) and then building up to multiple layers. Before proceeding to calculate higher order correlators, we briefly show that the time-ordered density-density correlator becomes trivial when averaged over temporal disorder (Section VI). Section VII presents the calculation of the time-ordered correlator $C_1$ for the NC-ST case, where we begin by establishing a vectorized notation to be used throughout the text. The calculation of $\overline{C_1}$ involves an average over two unitaries (Section VII S.1) and results in the TOC being given by a discrete random walk in 1D (Section VII S.2). Taking the continuum limit of the discrete evolution equation for $\overline{C_1}$ results in a diffusion equation in 1D (Section VII S.3) and a Gaussian broadening with standard deviation $\sigma(t) = \sqrt{2t}$. In Section VIII, the calculation of $\overline{C_2}$ proceeds analogously to that of $\overline{C_1}$ but is more involved not least because the disorder average is carried out over a product of four unitaries rather than two in the case of $\overline{C_1}$. The result is that $\overline{C_2}$ is the diagonal of a two-dimensional quantity whose discrete evolution equation, detailed in Section VIII S.2, is approximated by a 2D diffusion equation in the continuum limit, where the OTOC is approximated by a Gaussian with broadening width $\sigma(t) = \sqrt{t}$ (Section VIII S.3). After establishing the results for the NC-ST case, we extend these to the NC-T and C-ST cases by pointing out the differences with regards to NC-ST. The different structure of the NC-T circuit plays no role and both $\overline{C_1}$ (Section IX) and $\overline{C_2}$ (Section X) are exactly equal to the ones obtained for NC-ST. The same holds true for $\overline{C_1}$ in the C-ST case (Section XI). However, the conserving character of the unitaries used to build the circuit leads $\overline{C_2}$ to be different from its non-conserving counterpart. The details given in Section XII do not alter the qualitative behaviour in the continuum limit, affecting only the normalization of the Gaussian obtained before.

Lastly, the analytical results obtained for the three instances of free fermion evolution (NC-ST, NC-T and C-ST) are shown to agree with simulations and also with the continuum limit solutions (Section XIV). Furthermore, before concluding (Section XV) we present numerical results for a non particle conserving circuit with quenched spatial disorder (NC-S) that Anderson localizes (Section XIV S.1).

## II. FERMIONIC QUADRATIC HAMILTONIANS

Before proceeding to evaluate the TOC and OTOC, let us briefly introduce the system.

We consider free (or quadratic) fermions on a chain with $L$ sites. The Hamiltonian has, at most, quadratic terms in the fermionic operators, $a$ and $a^\dagger$, and it can be written as

$$\mathcal{H} = \frac{1}{2} A^\dagger H A \quad \text{with} \quad H = \begin{pmatrix} h & \Delta \\ \Delta^\dagger & -h^T \end{pmatrix}, \tag{S1}$$

where $A = (a_1, \cdots, a_L, a_1^\dagger, \cdots, a_L^\dagger)^T$ is the Nambu vector and $h = h^\dagger$ and $\Delta = -\Delta^T$. This system is invariant under

the exchange of creation and annihilation operators, i.e. it is particle-hole symmetric, which translates to

$$SHS = -H^T \quad \text{with} \quad S = \begin{pmatrix} 0 & \mathbb{1} \\ \mathbb{1} & 0 \end{pmatrix}. \tag{S2}$$

An operator $H$ respecting (S2) is said to be particle-hole (PH) symmetric.

Given the $2L \times 2L$ single-body Hamiltonian matrix, $H$, we may diagonalize the problem and compute arbitrary observables in terms of the single particle states. More specifically, having the two-point correlator $\chi = \langle AA^\dagger \rangle$, the mean value of any observable of the form $\mathcal{O} = \frac{1}{2} A^\dagger O A$ can be computed using

$$\langle \mathcal{O}(t) \rangle = \frac{1}{2} \text{tr} \left( \rho A^\dagger(t) O A(t) \right) = -\frac{1}{2} \text{tr} \left( O \chi(t) \right), \tag{S3}$$

with $\rho = e^{-\beta \mathcal{H}}/Z$ and $Z = \text{tr}(e^{-\beta \mathcal{H}})$ for a thermal density matrix (this is valid for our circuits with $\beta = 0$). Furthermore, we know that the fermionic operators $A(t)$ obey $\partial_t A(t) = -iHA(t)$, meaning that they evolve according to $A(t) = \exp(-iHt)A_0$. With this, and considering the initial $\chi = \langle AA^\dagger \rangle = \text{diag}(1-n_1, \ldots, 1-n_L, n_1, \ldots, n_L)$, we know how $\chi$ evolves:

$$\chi(t) = e^{-iHt} \chi e^{iHt}. \tag{S4}$$

This means that we have the tools to compute (S3). In particular, the entanglement entropy itself can be computed in a simple manner with [2]

$$S = -\text{tr}(\rho \log \rho) = -\text{tr}(\chi \log \chi). \tag{S5}$$

### III. SETTING UP THE RANDOM CIRCUITS

Unlike the gates used for generic dynamics, which needed only to respect unitarity, free fermion evolution operators are restricted to obey the fermionic anticanonical commutation relations.

In the single-particle basis, the Nambu vector $A_r = (a_r, a_{r+1}, a_r^\dagger, a_{r+1}^\dagger)^T$ evolves as $A_r(t + \Delta t) = u_{r,r+1} A_r(t)$, with $u_{r,r+1} = \exp(-iH\Delta t)$. This operator respects PH symmetry, i.e. $Su_{r,r+1}^T S = u_{r,r+1}^\dagger$, and it can be written as $u_{r,r+1} = V^\dagger \tilde{u}_{r,r+1} V$, with $\tilde{u}_{r,r+1}$ its representation in the Majorana basis, which respects $\tilde{u}_{r,r+1}^T \tilde{u}_{r,r+1} = \mathbb{1}$. Thus, the $4 \times 4$ unitaries $u$ can be obtained as

$$u = V^\dagger O V \quad \text{with} \quad V = \frac{1}{\sqrt{2}} \begin{pmatrix} \mathbb{1} & \mathbb{1} \\ -i\mathbb{1} & i\mathbb{1} \end{pmatrix} \tag{S6}$$

and $O$ a $4 \times 4$ random Haar-distributed orthogonal matrix. For the conserving case $u_{r,r+1} = v_{r,r+1} \oplus v_{r,r+1}^*$ (where $\oplus$ stands for direct sum), with $v_{r,r+1}$ a $2 \times 2$ Haar distributed unitary matrix.

The circuits used for the different instances of free fermion evolution studied, built using the gates just mentioned, are presented in Fig. S1.

### IV. TOC AND OTOC IN THE SINGLE-BODY PICTURE

The degree of non-commutativity of two local observables $\mathcal{W}$ and $\mathcal{V}$ centred, respectively, around positions 0 and $r$ with finite support is given by

$$\mathcal{C}(r,t) = \frac{1}{2} \langle |[\mathcal{W}(t), \mathcal{V}]|^2 \rangle = \mathcal{C}_1(r,t) - \mathcal{C}_2(r,t), \tag{S7}$$

where $\mathcal{W} = \mathcal{U}^\dagger(t) \mathcal{W} \mathcal{U}(t)$ and the TOC and OTOC are, respectively, $\mathcal{C}_1(r,t) = \langle \mathcal{W}^2(t) \mathcal{V}^2 \rangle$ and $\mathcal{C}_2(r,t) = \langle \mathcal{W}(t) \mathcal{V} \mathcal{W}(t) \mathcal{V} \rangle$. Quadratic many-body observables $\mathcal{V} = \mathcal{V}(0)$ and $\mathcal{W} = \mathcal{W}(0)$ can be written as $\mathcal{V} = 1/2 A^\dagger V A$ and $\mathcal{W} = 1/2 A^\dagger W A$, with $A = (a_1, \cdots, a_L, a_1^\dagger, \cdots, a_L^\dagger)^T$ the Nambu vector and $a_r$ and $a_r^\dagger$ the fermionic operators at position $r$. Considering that $\mathcal{W}$ and $\mathcal{V}$ have this form, we can express (S7) in terms of $V$ and $W$, the so called single-body observables. For this, it is convenient to write the relation between many and single-body operators

$$\mathcal{V} = \partial_v \Big|_{v=0} \exp\left(\frac{v}{2} A^\dagger V A\right) \quad \text{and} \quad \mathcal{W} = \partial_w \Big|_{w=0} \exp\left(\frac{w}{2} A^\dagger W A\right) \tag{S8}$$

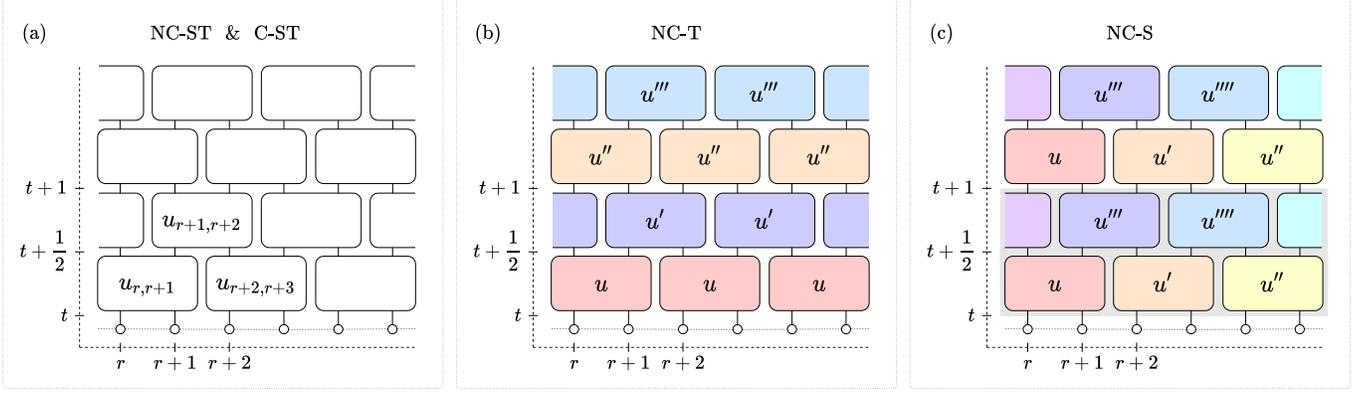

FIG. S1: (a) Circuit scheme for the NC-ST and C-ST cases in the $r-t$ plane: the two-site unitaries are applied at neighbouring sites in a brickwall pattern. The two cases differ in the gates used: the NC-ST case uses two-site free fermion gates $u_{r,r+1} = V^\dagger O V$, where $O$ is a $4 \times 4$ random Haar-distributed orthogonal matrix, while for C-ST $u_{r,r+1} = \mathrm{diag}(w,w)$, with $w$ a $2 \times 2$ random Haar-distributed unitary matrix. (b) Circuit scheme for the NC-T case. At each half time step, the same unitary $u$ is applied to all pairs of sites. (c) Circuit scheme for the NC-S case. One even and one odd layers are generated as usual, these are fixed and repeated in time (i.e. the grey box is repeated in time). In (b) and (c) the gates are free fermion gates, as for NC-ST.

This operator evolves as $\mathcal{W}(t) = \mathcal{U}^\dagger(t)\mathcal{W}\mathcal{U}(t)$, with

$$\mathcal{U}(t) = e^{-\frac{i}{2}A^\dagger H_t A} \ldots e^{-\frac{i}{2}A^\dagger H_1 A}. \tag{S9}$$

Replacing $\mathcal{V}$ and $\mathcal{W}$ by (S8) in (S7), $C_1(t)$ and $C_2(t)$ become

$$\mathcal{C}_k(r,t) = \partial_w\Big|_{w=0}\partial_v\Big|_{v=0}\partial_{w'}\Big|_{w'=0}\partial_{v'}\Big|_{v'=0}\tilde{C}_k(r,t) \qquad \text{with} \qquad j=1,2, \tag{S10}$$

$$\tilde{C}_1(r,t) = \frac{1}{2^L}\mathrm{tr}\left[\mathcal{U}^\dagger(t)e^{\frac{w+w'}{2}A^\dagger W A}\mathcal{U}(t)e^{\frac{v+v'}{2}A^\dagger W A}\right], \tag{S11}$$

$$\tilde{C}_2(r,t) = \frac{1}{2^L}\mathrm{tr}\left[\mathcal{U}^\dagger(t)e^{\frac{w}{2}A^\dagger W A}\mathcal{U}(t)e^{\frac{v}{2}A^\dagger V A}\mathcal{U}^\dagger(t)e^{\frac{w'}{2}A^\dagger W A}\mathcal{U}(t)e^{\frac{v'}{2}A^\dagger V A}\right] \tag{S12}$$

These expressions can be simplified with the properties,

$$e^{-\frac{1}{2}A^\dagger B_n A}\ldots e^{-\frac{1}{2}A^\dagger B_1 A} = e^{-\frac{1}{2}A^\dagger \tilde{B} A} \qquad \text{with} \qquad e^{\tilde{B}} = e^{B_1}\ldots e^{B_n}, \tag{S13}$$

$$\mathrm{tr}\left[e^{-\frac{1}{2}A^\dagger \tilde{B} A}\right] = \left[\det\left(1+e^{\tilde{B}}\right)\right]^{\frac{1}{2}}, \tag{S14}$$

$$\det(A) = \exp(\mathrm{tr}[\log A]), \tag{S15}$$

with the second one being valid for $\tilde{B}$ particle-hole symmetric. Applying these three expression to (S11) and (S12), we get

$$\tilde{C}_k(r,t) = \frac{1}{2^L}\mathrm{tr}\left[e^{-\frac{1}{2}A^\dagger \tilde{B}_k A}\right] = \frac{1}{2^L}\left[\det(1+e^{\tilde{B}_k})\right]^{\frac{1}{2}} = \frac{1}{2^L}\exp\left(\frac{1}{2}\mathrm{tr}\left[\log(1+e^{\tilde{B}_k})\right]\right), \tag{S16}$$

where $k=1,2$ and

$$e^{\tilde{B}_1} \equiv e^{(v+v')V}u_t^\dagger e^{(w+w')W}u_t, \tag{S17}$$

$$e^{\tilde{B}_2} \equiv e^{v'V}u_t^\dagger e^{w'W}u_t e^{vV}u_t^\dagger e^{wW}u_t, \tag{S18}$$

$$u_t \equiv e^{-iH_t}\ldots e^{-iH_1}. \tag{S19}$$

Finally, to obtain $\mathcal{C}(r,t)$ as a simple function of $W(t)$ and $V$, we must perform the derivatives in (S10), making use

of (S16). After doing so, and using $\operatorname{tr} \mathcal{V} = \operatorname{tr} \mathcal{W} = 0$ (which greatly simplifies the result), we obtain

$$\mathcal{C}_1(r,t) = \frac{1}{2^4} \left\{ \frac{1}{2} \operatorname{tr}[W(t)V]^2 + \frac{1}{4} \operatorname{tr}[W^2(t)] \operatorname{tr}[V^2] - \operatorname{tr}[W(t)VW(t)V] \right\}, \tag{S20}$$

$$\mathcal{C}_2(r,t) = \frac{1}{2^4} \left\{ \frac{1}{2} \operatorname{tr}[W(t)V]^2 + \frac{1}{4} \operatorname{tr}[W^2(t)] \operatorname{tr}[V^2] + \operatorname{tr}[W(t)VW(t)V] - 2 \operatorname{tr}[W^2(t)V^2] \right\}. \tag{S21}$$

Combining (S20) and (S21) in (S7) we obtain

$$C(r,t) = \frac{1}{2^3} \Big[ \mathcal{C}_1(r,t) - \mathcal{C}_2(r,t) \Big], \tag{S22}$$

with the new TOC and OTOC being, respectively,

$$C_1(r,t) = \operatorname{tr}[W^2(t)V^2], \tag{S23}$$
$$C_2(r,t) = \operatorname{tr}[W(t)VW(t)V]. \tag{S24}$$

These expressions are very similar to $\mathcal{C}_1(r,t)$ and $\mathcal{C}_2(r,t)$, with the many-body observables being replaced by the respective single-body ones. Note, however, that the correspondence between the many and single-body TOC and OTOC, given in (S20) and (S21), is not one-to-one.

The relations (S20) and (S21) can be simplified if we specify the observables to be the symmetrized particle number operator introduced below, i.e. $W = O_0$ and $V = O_r$. Since $O_r$ respects PH symmetry, one can show that $1/2 \operatorname{tr}[O_0(t)O_r]^2 = \operatorname{tr}[O_0(t)O_rO_0(t)O_r]$ and $\operatorname{tr}[O_0^2(t)] \operatorname{tr}[O_r^2] = 4$ such that

$$\mathcal{C}_1(r,t) = \frac{1}{2^4}, \tag{S25}$$

$$\mathcal{C}_2(r,t) = \frac{1}{2^4}[1 + 2C_2(r,t) - 2C_1(r,t)]. \tag{S26}$$

Knowing that the TOC and OTOC are given by (S23) and (S24) in terms of single-particle observables, we devote the next sections to computing these for the different types of circuits considered: NC-ST, NC-T and C-ST. We begin by carefully computing $\overline{C_1(x,t)}$ and $\overline{C_2(x,t)}$ for the NC-ST case. Then, we see that the results obtained apply to NC-T and C-ST with the exception of $\overline{C_2(x,t)}$ in the C-ST case. Before proceeding to do this, we address some useful topics. Namely, we introduce the observables to be used, commenting on their evolution operators, and some convenient notation for the single-particle states. We comment on the average of Haar-distributed orthogonal matrices, which will be needed when computing the TOC and OTOC for the NC-ST and NC-T cases, and of unitary matrices, needed for the C-ST case. Finally, and before proceeding to computing four-point correlators we see that the two-point correlator is trivially zero.

### 1. Symmetrized particle number observable

The ballistic / diffusive nature of the OTOC should not depend on the particular local observable used. For convenience, we consider the symmetrized particle number operator at position $r$, i.e.

$$\mathcal{O}_r(0) = \mathcal{O}_r = \frac{1}{2}(a_r^\dagger a_r - a_r a_r^\dagger) = a_r^\dagger a_r - 1/2 \quad \Leftrightarrow \quad O_r(0) = O_r = \operatorname{diag}(\ldots, 1_r, \ldots, -1_{r+L}, \ldots). \tag{S27}$$

Specifically, $\mathcal{V}$ and $\mathcal{W}$ are the symmetrized number operators initially localized at positions $r$ and $0$, respectively: $\mathcal{V} = \mathcal{O}_r \Leftrightarrow V = O_r$ and $\mathcal{W} = \mathcal{O}_0 \Leftrightarrow W = O_0$.

### 2. Single-particle states

Rewriting our observables using Nambu's notation as $\mathcal{O} = 1/2 A^\dagger O A$ allows us to work with the single-body observable $O$ in the single-body basis which includes a total of $2L$ single-particle states. A single-particle state corresponds to a site $r$ being occupied or not, i.e. we have either a particle or a hole identified, respectively, by indices $p$ and $h$: $|r,p\rangle = |1_r\rangle$ or $|r,h\rangle = |0_r\rangle$. It is convenient for us to label these states by

$$|\alpha\rangle = |r_\alpha, s_\alpha\rangle = |2R_{r_\alpha} + b_{r_\alpha}, s_\alpha\rangle, \tag{S28}$$

where $r_\alpha = 0, ..., L-1$ runs over the lattice sites and $s_\alpha \in \{p, h\}$ is the particle-hole index. In the second equality the position index $r_\alpha$ is decomposed as $r_\alpha = 2R_{r_\alpha} + b_{r_\alpha}$, where $R_{r_\alpha} = 0, ..., L/2-1$ labels the pair of sites $(r_\alpha, r_\alpha + 1)$ and $b_{r_\alpha} \in \{0, 1\}$ labels the position within the pair $R_{r_\alpha}$. This decomposition will prove useful since it emphasizes the structure of the circuit built out of two-site unitaries. Also, we use $|\bar{\alpha}\rangle$ to refer to $|r_\alpha, \bar{s}_\alpha\rangle$ where $\bar{p} = h$ and $\bar{h} = p$. For convenience, we will use the different notations interchangeably.

### 3. Single-body time evolution operators

In the single-body picture, each observable $O_r$ evolves according to $O_r(t) = U^\dagger(t) O_r U(t)$. Each $U(t)$ is composed of $t$ gates $U = U_{\text{even}} U_{\text{odd}}$ composed of one even and one odd layer. Establishing that $U_{r,r+1}$ is the $2L \times 2L$ operator which acts non-trivially on sites $r$ and $r+1$, an even / odd layer can be written as

$$U_{\text{even}} = \prod_{R=0}^{L/2-1} U_{2R,2R+1} \quad \text{and} \quad U_{\text{odd}} = \prod_{R=0}^{L/2-1} U_{2R-1,2R}, \tag{S29}$$

where we label the position using the pair index introduced in (S28) and where the sites are implicitly acted upon by modulo $L$, due to periodic boundary conditions (this will be omitted in the future). In its turn, $U_{r,r+1} = u_{r,r+1} + \bar{\mathbb{P}}_{r,r+1}$, where $u_{r,r+1} = \sum_{s,s'} \sum_{x,x' \in \{r,r+1\}} |x,s\rangle \langle x,s| u |x',s'\rangle \langle x',s'|$ is the $4 \times 4$ matrix responsible for mixing the pair $(r, r+1)$, while $\bar{\mathbb{P}}_{r,r+1} = \sum_s \sum_{x \notin \{r,r'\}} |x,s\rangle \langle x,s|$ is the trivial projector into the complement of the pair. Taking this into account, we can write each layer in terms of the non-trivial $u$s as

$$U_{\text{even}} = \sum_{R=0}^{L/2-1} u_{2R,2R+1} \quad \text{and} \quad U_{\text{odd}} = \sum_{R=0}^{L/2-1} u_{2R-1,2R}. \tag{S30}$$

This identity will prove to be useful.

## V. AVERAGE OF PRODUCTS OF HAAR-DISTRIBUTED ORTHOGONAL AND UNITARY MATRICES

The gates which compose our circuits are random and distributed according to the Haar measure. This is the unique measure invariant under group multiplication, weighting different regions of the probability space equally and thus behaving like a uniform distribution [3]. The average with respect to the Haar probability measure $\mu$ on the matrix probability state $\text{M}(N)$ is denoted by $\int_{\text{M}(N)} \ldots d\mu(M)$ and abbreviated as a line over the averaged quantity. We will address to the average of products of entries of some Haar-distributed gate $u$ as moments or matrix integrals of $u$.

Since non-conserving free fermion gates are obtained from real orthogonal matrices $O$ according to $u = VOV^\dagger$, their moments can be obtained from those of Haar-distributed orthogonal matrices. The average of products of matrix elements of a single orthogonal matrix $O$ with respect to the Haar probability measure $\mu$ on $\text{O}(N)$ is obtained in Refs. [4]. In particular, it is given by the Corollary 3.4. in [4], which uses the so called Weingarten function on the permutation group defined in Theorem 3.9 and given explicitly in Sec. 6. In particular, we are interested in the products of two and four entries:

$$\overline{\langle \alpha_1| O |\beta_1\rangle \langle \alpha_2| O |\beta_2\rangle} = \frac{1}{N} \delta_{\alpha_1 \alpha_2} \delta_{\beta_1 \beta_2}, \tag{S31}$$

$$\overline{\langle \beta_1| O |\alpha_1\rangle \langle \beta_2| O |\alpha_2\rangle \langle \beta_3| O |\alpha_3\rangle \langle \beta_4| O |\alpha_4\rangle}$$
$$= \frac{1}{N(N-1)(N+2)} \Bigg\{ \delta_{\beta_1 \beta_4} \delta_{\beta_2 \beta_3} \Big[ (N+1) \delta_{\alpha_1 \alpha_4} \delta_{\alpha_2 \alpha_3} - \delta_{\alpha_1 \alpha_2} \delta_{\alpha_3 \alpha_4} - \delta_{\alpha_1 \alpha_3} \delta_{\alpha_2 \alpha_4} \Big]$$
$$+ \delta_{\beta_1 \beta_2} \delta_{\beta_3 \beta_4} \Big[ -\delta_{\alpha_1 \alpha_4} \delta_{\alpha_2 \alpha_3} + (N+1) \delta_{\alpha_1 \alpha_2} \delta_{\alpha_3 \alpha_4} - \delta_{\alpha_1 \alpha_3} \delta_{\alpha_2 \alpha_4} \Big]$$
$$+ \delta_{\beta_1 \beta_3} \delta_{\beta_2 \beta_4} \Big[ -\delta_{\alpha_1 \alpha_4} \delta_{\alpha_2 \alpha_3} - \delta_{\alpha_1 \alpha_2} \delta_{\alpha_3 \alpha_4} + (N+1) \delta_{\alpha_1 \alpha_3} \delta_{\alpha_2 \alpha_4} \Big] \Bigg\}, . \tag{S32}$$

Equivalent expressions for free fermion gates are obtained by doing $u = VOV^\dagger$, being given by (S40), (S62) and (S63). These are very similar to the ones above, with the difference that indices of the type $\bar{\alpha}$ appear.

For the specific particle conserving case, the average over free fermion gates $u = \text{diag}(w, w)$, where $w$ are unitary gates, can be obtained from the moments of unitary matrices, which are given by the Corollary 2.4 in [4]. In particular, the second moment of a $N \times N$ unitary matrix $w$, $\overline{\langle \alpha_1 | w | \beta_1 \rangle^* \langle \alpha_2 | w | \beta_2 \rangle}$, is equal to (S31) while the fourth moment is given by (S109).

## VI. TWO-POINT CORRELATOR

With the symmetrized particle number operator as our observable, it is natural to consider $\text{tr}[\overline{O_0(t)O_x}]$ as the two-point correlator. Its dynamics is determined by the evolution of $O_0(t+1/2) = U^\dagger O_0(t) U$, where $U$ alternates between $U_{\text{even}}$ and $U_{\text{odd}}$, given by (S30). This can be averaged considering $\overline{\langle \alpha_1 | u | \beta_1 \rangle^* \langle \alpha_2 | u | \beta_2 \rangle} = 1/N \delta_{\alpha_1 \alpha_2} \delta_{\beta_1 \beta_2}$, becoming

$$\overline{O_0(t+1/2)} = \sum_R \sum_{r,r' \in \{2R, 2R+1\}} \sum_{s,s'} |r', s'\rangle \langle r, s | \overline{O_0(t)} | r, s \rangle \langle r', s' |, \tag{S33}$$

where we considered $U = U_{\text{even}}$, with analogous results following for $U = U_{\text{odd}}$. Since $\sum_s \langle r, s | \overline{O_0(t)} | r, s \rangle = 0$ (due to PH symmetry), it follows that $\text{tr}[\overline{O_0(t)O_x}] = 0$. This leads us to look for non-trivial behaviour in higher order correlators.

## VII. NC-ST: AVERAGED TOC, $\overline{C_1(x,t)}$

We start by considering the spatio-temporal noisy free fermion circuit without particle conservation (NC-ST), drawn in Fig. S1 (a). We first compute $\overline{C_1(r,t)}$ and then $\overline{C_2(r,t)}$. The dynamics of the later will prove to be more intricate and so we employ a vectorized notation which helps clarifying its behaviour. For a matter of consistency, this is also applied to $\overline{C_1(r,t)}$. Accordingly, we consider $C_1(r,t)$ as the overlap of two vectorized operators

$$C_1(r,t) = \text{tr}[O_0^2(t) O_r^2] = \langle\langle O_r^2 || O_0^2(t) \rangle\rangle, \tag{S34}$$

with $||O_r^2(t)\rangle\rangle$ given by

$$||O_r^2(t)\rangle\rangle = \sum_{x,x'=0}^{L-1} \sum_{s,s'} ||x, s; x', s'\rangle\rangle \langle x, s | O_r^2(t) | x', s' \rangle, \tag{S35}$$

where $||x, s; x', s'\rangle\rangle = |x, s\rangle \otimes \langle x', s'|^T$ and the particle-hole index $s$ sums over $p$ and $h$ (we usually omit these). Also, with $O_r$ the symmetrized number operator at position $r$ given by (S27) we have

$$||O_r^2\rangle\rangle = \sum_s ||r, s; r, s\rangle\rangle. \tag{S36}$$

Having established this vectorized formalism as our framework, we proceed to average $C_1(r,t)$ over multiple realizations of the random circuit and we evaluate the ensuing dynamics.

### S.1. Time evolution of $||\overline{O_0^2(t)}\rangle\rangle$

The dynamics of $C_1(r,t)$ is completely contained in $||O_0^2(t)\rangle\rangle$. The operator $O_0^2(t)$ evolves as $O_0^2(t+1) = U^\dagger O_0^2(t) U$ which, in the vectorized notation, translates to

$$||O_0^2(t+1)\rangle\rangle = U^\dagger \otimes U^T ||O_0^2(t)\rangle\rangle. \tag{S37}$$

Each circuit is a succession of layers of two-site gates acting on even and odd pairs of sites. In one unit of time one even and one odd layers are applied, $U = U_{\text{odd}} U_{\text{even}}$, such that

$$||O_0^2(t+1)\rangle\rangle = \left(U^\dagger \otimes U^T\right)_{\text{even}} \left(U^\dagger \otimes U^T\right)_{\text{odd}} ||O_0^2(t)\rangle\rangle$$

$$= \left(\sum_{R,R'=0}^{L/2-1} u^\dagger_{2R,2R+1} \otimes u^T_{2R',2R'+1}\right) \left(\sum_{R,R'=0}^{L/2-1} u^\dagger_{2R-1,2R} \otimes u^T_{2R'-1,2R'}\right) ||O_0^2(t)\rangle\rangle, \tag{S38}$$

where the sites are labelled using the pair indices and where we used (S30).

### 1. Average over disorder realizations

Next, we average random realizations of the circuit to obtain the leading order behaviour of $C_1(r,t)$. The average $||O_0^2(t)\rangle\rangle$ reduces to the average over the two-site gates composing the circuit. Since these are randomly chosen, they are uncorrelated and we can perform the average independently at different half-time steps. Thus, we only need to find

$$\overline{u_{r,r+1}^\dagger \otimes u_{r',r'+1}^T} = ||\alpha_1\alpha_2\rangle\rangle \, \overline{\langle\beta_1|\, u_{r,r+1}\,|\alpha_1\rangle^* \langle\beta_2|\, u_{r',r'+1}\,|\alpha_2\rangle} \, \langle\langle\beta_1\beta_2||. \tag{S39}$$

This demands that we compute the average of products of entries of some Haar-distributed free fermion gate $u$, also called the moments or matrix integrals of $u$. A short discussion on the topic was presented in Sec. V. Summarizing, expressions for the moments of $u$ can be obtained from equivalent expressions existing for an Haar-distributed orthogonal matrix $O$ and only moments of an even number of entries of $u$ are non-zero. This implies $\overline{u_{r,r+1}^\dagger \otimes u_{r',r'+1}^T} = \delta_{r,r'} \overline{u_{r,r+1}^\dagger \otimes u_{r,r+1}^T}$. Considering that the second moment of some free fermion gate $u$ is given by (see Sec. V)

$$\overline{\langle\alpha_1|\,u\,|\beta_1\rangle^* \langle\alpha_2|\,u\,|\beta_2\rangle} = \frac{1}{N}\delta_{\alpha_1\alpha_2}\delta_{\beta_1\beta_2}, \tag{S40}$$

where $N = \mathrm{rank}(u) = 4$, we get

$$\overline{u_{r,r+1}^\dagger \otimes u_{r,r+1}^T} = \frac{1}{N}||\alpha\alpha\rangle\rangle \langle\langle\beta\beta|| = ||\phi_r\rangle\rangle \langle\langle\phi_r||, \tag{S41}$$

i.e. the projector evolving the pair of sites $(r, r+1)$ from some time $t$ to $t + 1/2$, with

$$||\phi_r\rangle\rangle = \frac{1}{2} \sum_{x\in\{r,r+1\}} \sum_s ||x,s;x,s\rangle\rangle, \tag{S42}$$

which obeys $\langle\langle\phi_r||\phi_{r'}\rangle\rangle = \delta_{r',r} + \frac{1}{2}\delta_{r',r\pm 1}$. Also, note that $||\phi_r\rangle\rangle \langle\langle\phi_r||$ leads $\overline{O_0^2(t)}$ to be always diagonal. Plugging these results into the averaged equivalent of (S38) leads to

$$||\overline{O_0^2(t+1)}\rangle\rangle = \left(\sum_{R=0}^{L/2-1} ||\phi_{2R}\rangle\rangle \langle\langle\phi_{2R}||\right) \left(\sum_{R=0}^{L/2-1} ||\phi_{2R-1}\rangle\rangle \langle\langle\phi_{2R-1}||\right) ||\overline{O_0^2(t)}\rangle\rangle, \tag{S43}$$

To know $||\overline{O_0^2(t)}\rangle\rangle$ at any $t$ we just need to start from the initial condition $||\overline{O_0^2(0)}\rangle\rangle = \sum_s ||0,s;0,s\rangle\rangle$ and apply the above expression successively. Next, we do this and simplify (S43).

### 2. Dynamics of $||\overline{O_0^2(t)}\rangle\rangle$

Starting with $||\overline{O_0^2(0)}\rangle\rangle = \sum_s ||0,s;0,s\rangle\rangle$ and applying the first odd layer of the circuit we obtain

$$||\overline{O_0^2(t=1/2)}\rangle\rangle = \left(\sum_{R=0}^{L/2-1} ||\phi_{2R-1}\rangle\rangle \langle\langle\phi_{2R-1}||\right) ||\overline{O_0^2(0)}\rangle\rangle = ||\phi_{-1}\rangle\rangle \langle\langle\phi_{-1}||\overline{O_0^2(0)}\rangle\rangle = ||\phi_{-1}\rangle\rangle, \tag{S44}$$

to which we can apply a second even layer to get

$$\begin{aligned}||\overline{O_0^2(t=1)}\rangle\rangle &= \left(\sum_{R=0}^{L/2-1} ||\phi_{2R}\rangle\rangle \langle\langle\phi_{2R}||\right) ||\overline{O_0^2(1/2)}\rangle\rangle \\ &= \Big(||\phi_{-2}\rangle\rangle \langle\langle\phi_{-2}|| + ||\phi_0\rangle\rangle \langle\langle\phi_0||\Big) ||\phi_{-1}\rangle\rangle = \frac{1}{2}\Big(||\phi_{-2}\rangle\rangle + ||\phi_0\rangle\rangle\Big).\end{aligned} \tag{S45}$$

Also, mixing between neighbouring pairs in successive layers is mediated by

$$||\phi_r\rangle\rangle \langle\langle\phi_r||\phi_{r\pm 1}\rangle\rangle = \frac{1}{2}||\phi_r\rangle\rangle, \tag{S46}$$

which guarantees that the subspace $\{||\phi_r\rangle\rangle\}$ is closed under time evolution. This, allied with (S45), allows the decomposition

$$||\overline{O_0^2(t)}\rangle\rangle = \begin{cases} \sum_{R=0}^{L/2-1} ||\phi_{2R}\rangle\rangle \langle\langle\phi_{2R}||\overline{O_0^2(t)}\rangle\rangle & , \text{ for } t \text{ integer} \\ \sum_{R=0}^{L/2-1} ||\phi_{2R-1}\rangle\rangle \langle\langle\phi_{2R-1}||\overline{O_0^2(t)}\rangle\rangle & , \text{ for } t \text{ half-integer} \end{cases} \quad (S47)$$

i.e. $\{||\phi_{2R}\rangle\rangle \mid R = 0, \ldots, L/2 - 1\}$ and $\{||\phi_{2R-1}\rangle\rangle \mid R = 0, \ldots, L/2 - 1\}$ for $t$ integer and half-integer, respectively, are the natural orthonormal basis for our problem. This is, they take full advantage of the structure coming from the average of gates, meaning that $\langle\langle r, s; r, s||\overline{O_0^2(t)}\rangle\rangle = \langle\langle r+1, s; r+1, s||\overline{O_0^2(t)}\rangle\rangle$, where $(r, r+1)$ forms a pair, and of the PH symmetry of $O_0(t)$, which leads to $\langle\langle r, p; r, p||\overline{O_0^2(t)}\rangle\rangle = \langle\langle r, h; r, h||\overline{O_0^2(t)}\rangle\rangle$. Having established the good basis to use, we can employ (S46) to obtain, for $t \geq 1/2$,

$$\langle\langle\phi_r||\overline{O_0^2(t+1/2)}\rangle\rangle = \langle\langle\phi_r|| \left( ||\phi_{r-1}\rangle\rangle \langle\langle\phi_{r-1}|| + ||\phi_{r+1}\rangle\rangle \langle\langle\phi_{r+1}|| \right) ||\overline{O_0^2(t)}\rangle\rangle$$
$$= \frac{1}{2} \left( \langle\langle\phi_{r-1}|| + \langle\langle\phi_{r+1}|| \right) ||\overline{O_0^2(t)}\rangle\rangle, \quad (S48)$$

i.e. the time-evolution of $\langle\langle\phi_r||\overline{O_0^2(t)}\rangle\rangle$ can be understood as an averaging process with contributions from the nearest neighbours. Applying this twice gives

$$\langle\langle\phi_r||\overline{O_0^2(t+1)}\rangle\rangle = \frac{1}{4} \left( \langle\langle\phi_{r-2}|| + 2\langle\langle\phi_r|| + \langle\langle\phi_{r+2}|| \right) ||\overline{O_0^2(t)}\rangle\rangle. \quad (S49)$$

This recursive expression associated to $||\overline{O_0^2(t)}\rangle\rangle = \sum_{R=0}^{L/2-1} ||\phi_{2R}\rangle\rangle \langle\langle\phi_{2R}||\overline{O_0^2(t)}\rangle\rangle$ is equivalent to (S43), but more transparent than the latter.

### S.2. Prescription to obtain $\overline{C_1(r,t)}$

Now that we have reduced $||\overline{O_0^2(t)}\rangle\rangle$ to its core, we focus back on the TOC.

For $t = 0$, we use the initial condition $||\overline{O_0^2(0)}\rangle\rangle = \sum_s ||0, s; 0, s\rangle\rangle$ in (S34) to obtain

$$\overline{C_1(r,0)} = C_1(r,0) = \langle\langle O_r^2||O_0^2(0)\rangle\rangle = 2\delta_{r,0}. \quad (S50)$$

For $t \geq 1$, we can decompose $||\overline{O_0^2(t)}\rangle\rangle$ using (S47) to obtain

$$\overline{C_1(r,t)} = \langle\langle O_r^2||\overline{O_0^2(t)}\rangle\rangle = \langle\langle O_r^2|| \sum_{R=0}^{L/2-1} ||\phi_{2R}\rangle\rangle \langle\langle\phi_{2R}||\overline{O_0^2(t)}\rangle\rangle = \sum_{R=0}^{L/2-1} (\delta_{r,2R} + \delta_{r,2R+1}) \langle\langle\phi_{2R}||\overline{O_0^2(t)}\rangle\rangle, \quad (S51)$$

i.e. $\overline{C_1(2R,t)} = \overline{C_1(2R+1,t)} = \langle\langle\phi_{2R}||\overline{O_0^2(t)}\rangle\rangle$. The dynamics of $\overline{C_1(r,t)}$ is then determined by that of $\langle\langle\phi_{2R}||\overline{O_0^2(t)}\rangle\rangle$, which we just saw is given by (S49), starting with (S45) as the initial condition, i.e.

$$\overline{C_1(2R, t+1)} = \overline{C_1(2R+1, t+1)} = \begin{cases} \frac{1}{2}(\delta_{R-1,0} + \delta_{R,0}) & , \text{ for } t = 0 \\ \frac{1}{4} \left( \overline{C_1(2R-2,t)} + 2\overline{C_1(2R,t)} + \overline{C_1(2R+2,t)} \right) & , \text{ for } t \geq 1 \end{cases}. \quad (S52)$$

Having broken down the TOC given initially by (S34), we end up with a very clean picture: the circuit's structure leads $\overline{C_1(r,t)}$ to depend only on the pair $R$ to which $r$ belongs and evolving it comes down to performing a weighted average as specified in (S52). This averaging process is represented pictorially in Fig. S2. In the panel (c) the 'brickwall' structure of the circuit ensures that, as time evolves, $\overline{C_1(r,t)}$ spreads across the system. Although the boundary in the $r - t$ plane between the region with zero and non-zero $\overline{C_1(r,t)}$ describes a light cone, the weights are concentrated around $r = 0$ such that the process is diffusive and not ballistic. Next, we argue in favour of this diffusive behaviour by going to the continuum limit.

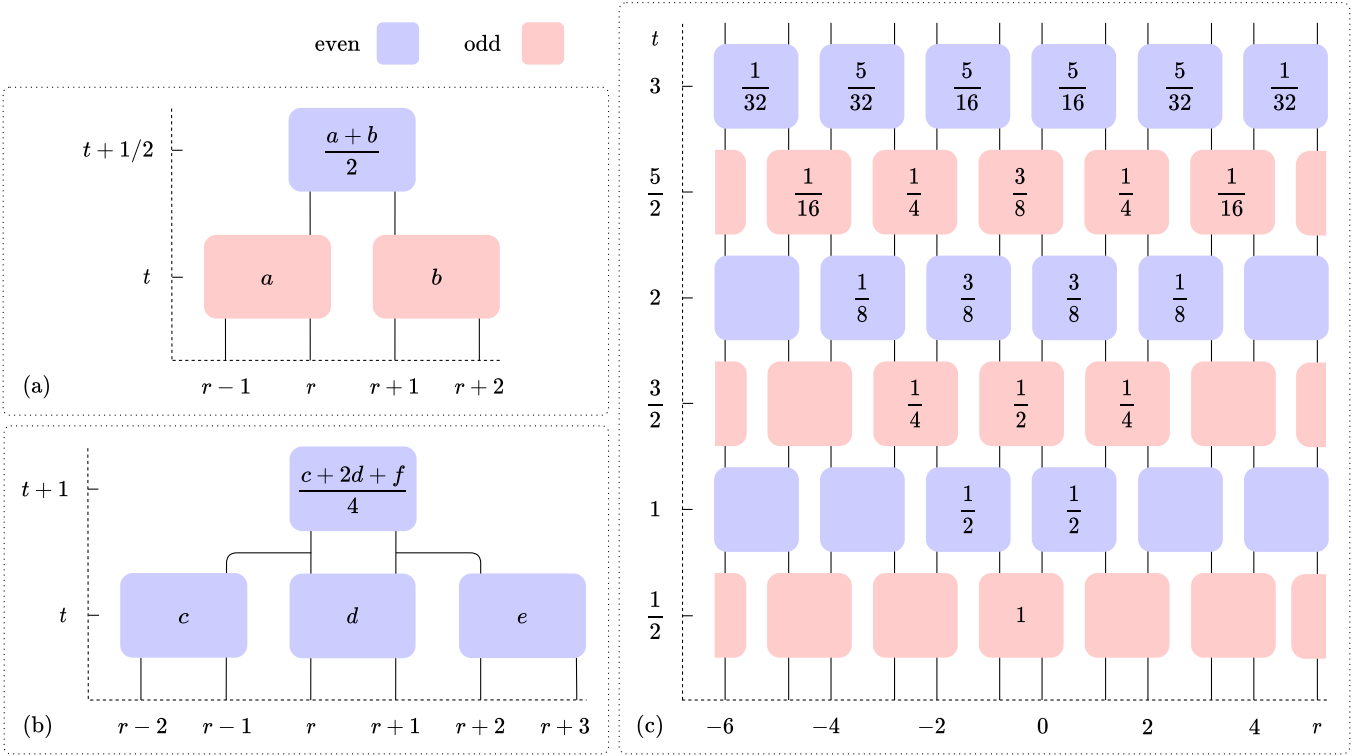

FIG. S2: The values in the boxes are $\overline{C_1(r,t)}$ in the $r-t$ plane (in empty boxes $\overline{C_1(r,t)} = 0$). Panels (a) and (b) translate Eqs. (S48) and (S49) to a schematic, representing the evolution of $\overline{C_1(r,t)}$ in one half and one full time steps. The scheme in (b) is obtained by applying (a) twice. (c) The previous schemes can be applied to obtain $\overline{C_1(r,t)}$, starting with $C_1(r,0) = 2\delta_{r,0}$ as the initial condition. Even layers are depicted in blue while odd layers appear in red.

### S.3. Continuum limit of $\overline{C_1(r,t)}$

The OTOC is described by (S52) for $t \geq 1$. We show that this is a discrete diffusion equation in 1D by recovering the continuum diffusion equation from it.

We consider the scaling form $aC'_1(x = ra, \tau = ta^2)$. For $a = 1$, this coincides with the discrete $C_1(r,t)$. If we let $a \to 0$, this approximates $C_1(r,t)$ in the continuum limit. After making this identification in (S52), where $r = 2R$, we Taylor expand it up to $\mathcal{O}(a^3)$, obtaining the one-dimensional continuum diffusion equation

$$\partial_\tau \overline{C'_1(x,\tau)} = D_1 \partial_x^2 \overline{C'_1(x,\tau)}, \tag{S53}$$

with diffusion coefficient $D_1 = 1$. Using the initial condition (S50), which translates to $C'_1(x, t = 0) = 2\delta(x)$, the solution to (S53) approximating $\overline{C_1(r,t)}$ for large $t$ and $L$ is

$$\overline{C_1(2R,t)} = \overline{C_1(2R+1,t)} \simeq a\overline{C'_1(x = 2Ra, \tau = ta^2)} = \frac{A}{\sqrt{2\pi}\sigma(t)} \exp\left(-\frac{(2R)^2}{2\sigma(t)^2}\right), \tag{S54}$$

i.e. a Gaussian normalized to $A = 2$ with standard deviation $\sigma = \sqrt{2t}$. Having recovered the 1D diffusion equation (S53) shows that, for large enough times, the exact (S52) diffuses in the $r-t$ plane.

### VIII. NC-ST: AVERAGED OTOC, $\overline{C_2(r,t)}$

We move on to calculate the OTOC itself for the non-conserving circuit with randomness in space and time (NC-ST). Although this will prove to be more intricate than $\overline{C_1(r,t)}$, the procedure follows the same steps.

We start by writing $C_2(r,t)$ as the overlap of two vectorized operators

$$C_2(r,t) = \text{tr}\left[O_0(t)O_r O_0(t)O_r\right] = \langle\langle Q_r|| S ||Q_0(t)\rangle\rangle, \tag{S55}$$

with $Q_r \equiv O_r \otimes O_r$ and

$$||Q_r(t)\rangle\rangle = \sum_{\alpha\beta\mu\nu} ||\alpha\beta\mu\nu\rangle\rangle \langle\alpha\beta|Q_r(t)|\mu\nu\rangle, \tag{S56}$$

$$S = \sum_{\alpha\beta\mu\nu} ||\alpha\beta\mu\nu\rangle\rangle \langle\langle\alpha\beta\nu\mu||, \tag{S57}$$

where $||\alpha\beta\mu\nu\rangle\rangle = |\alpha\beta\rangle \otimes \langle\mu\nu|^T$ and the operator $S$ reorders the indices such that $\langle\langle Q_r|| S ||Q_0(t)\rangle\rangle$ is the trace present in (S55). Also, with $O_r$ the symmetrized number operator at position $r$, we have

$$||Q_r\rangle\rangle = \sum_s ||rs, rs, rs, rs\rangle\rangle - ||rs, r\bar{s}, rs, r\bar{s}\rangle\rangle. \tag{S58}$$

We observe that in the vectorized notation a parallel is established between $C_1(r,t) = \langle\langle O_r^2|| O_0^2(t)\rangle\rangle$ and $C_2(r,t) = \langle\langle Q_r|| S ||Q_0(t)\rangle\rangle$: $||O_r^2\rangle\rangle$ and $||O_0^2(t)\rangle\rangle$ are to $C_1(r,t)$ as $||Q_r\rangle\rangle$ and $||Q_0(t)\rangle\rangle$ are to $C_2(r,t)$. The dynamics of the latter is set by $||Q_0(t)\rangle\rangle$, which we proceed to evaluate.

### S.1. Time evolution of $||\overline{Q_0(t)}\rangle\rangle$

The dynamics of $C_2(r,t)$ is determined by that of $||Q_0(t)\rangle\rangle$. As done for $C_1(r,t)$, we first compute how $||Q_0(t)\rangle\rangle$ evolves and then we average the result obtained.

From $O_0(t+1) = U^\dagger O_0(t) U$ it follows that $Q_0(t+1) = (U^\dagger \otimes U^\dagger) Q_0(t) (U \otimes U)$ which, in the vectorized notation, translates to

$$||Q_0(t+1)\rangle\rangle = U^\dagger \otimes U^\dagger \otimes U^T \otimes U^T ||Q_0(t)\rangle\rangle. \tag{S59}$$

In one time step two layers of the circuit are applied such that $U = U_{\text{odd}} U_{\text{even}}$. Using (S30), it follows that

$$\begin{aligned}
||Q_0(t+1)\rangle\rangle &= (U^\dagger \otimes U^\dagger \otimes U^T \otimes U^T)_{\text{even}} (U^\dagger \otimes U^\dagger \otimes U^T \otimes U^T)_{\text{odd}} ||Q_0(t)\rangle\rangle \\
&= \left(\sum_{R,R',R'',R'''=0}^{L/2-1} u^\dagger_{2R,2R+1} \otimes u^\dagger_{2R',2R'+1} \otimes u^T_{2R'',2R''+1} \otimes u^T_{2R''',2R'''+1}\right) \\
&\quad \left(\sum_{R,R',R'',R'''=0}^{L/2-1} u^\dagger_{2R-1,2R} \otimes u^\dagger_{2R'-1,2R'} \otimes u^T_{2R''-1,2R''} \otimes u^T_{2R'''-1,2R'''}\right) ||Q_0(t)\rangle\rangle,
\end{aligned} \tag{S60}$$

valid for some realization of the circuit.

#### 1. Average over disorder realizations

Wishing to obtain $||\overline{Q_0(t)}\rangle\rangle$, we move on to average (S60) over random realizations of the circuit. As before, we can take the average of each layer independently from the others. Thus, we wish to obtain

$$\overline{u^\dagger_{r,r+1} \otimes u^\dagger_{r',r'+1} \otimes u^T_{r'',r''+1} \otimes u^T_{r''',r'''+1}} = \\
||\alpha_1\alpha_2\alpha_3\alpha_4\rangle\rangle \overline{\langle\beta_1|u_{r,r+1}|\alpha_1\rangle^* \langle\beta_2|u_{r',r'+1}|\alpha_2\rangle^* \langle\beta_3|u_{r'',r''+1}|\alpha_3\rangle \langle\beta_4|u_{r''',r'''+1}|\alpha_4\rangle} \langle\langle\beta_1\beta_2\beta_3\beta_4||. \tag{S61}$$

To perform this average of a product of entries of $u$ we refer again to Sec. V. Remember that only even moments of $u$ are non-zero. The second moment is given either by (S40) or

$$\overline{\langle\beta_1|u|\alpha_1\rangle \langle\beta_2|u|\alpha_2\rangle} = \overline{\langle\beta_1|u|\alpha_1\rangle^* \langle\beta_2|u|\alpha_2\rangle^*} = \frac{1}{N}\delta_{\bar{\alpha}_1\alpha_2}\delta_{\bar{\beta}_1\beta_2} \tag{S62}$$

and the fourth moment is given by

$$\overline{\langle\beta_1|\,u\,|\alpha_1\rangle^*\,\langle\beta_2|\,u\,|\alpha_2\rangle^*\,\langle\beta_3|\,u\,|\alpha_3\rangle\,\langle\beta_4|\,u\,|\alpha_4\rangle}$$
$$=\frac{1}{N(N-1)(N+2)}\bigg\{\delta_{\beta_1\beta_4}\delta_{\beta_2\beta_3}\Big[(N+1)\delta_{\alpha_1\alpha_4}\delta_{\alpha_2\alpha_3}-\delta_{\bar{\alpha}_1\alpha_2}\delta_{\bar{\alpha}_3\alpha_4}-\delta_{\alpha_1\alpha_3}\delta_{\alpha_2\alpha_4}\Big]$$
$$+\delta_{\bar{\beta}_1\beta_2}\delta_{\bar{\beta}_3\beta_4}\Big[-\delta_{\alpha_1\alpha_4}\delta_{\alpha_2\alpha_3}+(N+1)\delta_{\bar{\alpha}_1\alpha_2}\delta_{\bar{\alpha}_3\alpha_4}-\delta_{\alpha_1\alpha_3}\delta_{\alpha_2\alpha_4}\Big]$$
$$+\delta_{\beta_1\beta_3}\delta_{\beta_2\beta_4}\Big[-\delta_{\alpha_1\alpha_4}\delta_{\alpha_2\alpha_3}-\delta_{\bar{\alpha}_1\alpha_2}\delta_{\bar{\alpha}_3\alpha_4}+(N+1)\delta_{\alpha_1\alpha_3}\delta_{\alpha_2\alpha_4}\Big]\bigg\},\tag{S63}$$

with $N = \mathrm{rank}(u) = 4$. This means that, when performing the average in (S61), either two or four entries of $u$ must refer to the same unitary, leaving us with four possible terms. The case where all entries refer to the same unitary (i.e. $r = r' = r'' = r'''$) contributes with one term while the other three come from two different unitaries appearing, for which there are three configurations corresponding to the pairings we can make out of four elements: $(12)(34)$, $(13)(24)$ and $(14)(23)$. Taking (S40), (S62) and (S63) into account, the sum over (S61) becomes

$$\sum_{r,r',r'',r'''=0}^{L/2-1}\overline{u^\dagger_{r,r+1}\otimes u^\dagger_{r',r'+1}\otimes u^T_{r'',r''+1}\otimes u^T_{r''',r'''+1}}=\sum_{r,r'=0}^{L/2-1}v_{r,r'},\tag{S64}$$

with

$$v_{r,r'} = (1-\delta_{rr'})\Big(\,||\Theta\rangle\rangle\,\langle\langle\Theta||+||\bar{\Theta}\rangle\rangle\,\langle\langle\bar{\Theta}||+||\tilde{\Theta}\rangle\rangle\,\langle\langle\tilde{\Theta}||\,\Big)_{r,r'}$$
$$+\frac{\delta_{rr'}}{(N+2)}\bigg(\Big[(N+1)\,||\Theta\rangle\rangle-||\bar{\Theta}\rangle\rangle-||\tilde{\Theta}\rangle\rangle\Big]\langle\langle\Theta||$$
$$+\Big[-||\Theta\rangle\rangle+(N+1)\,||\bar{\Theta}\rangle\rangle-||\tilde{\Theta}\rangle\rangle\Big]\langle\langle\bar{\Theta}||+\Big[-||\Theta\rangle\rangle-||\bar{\Theta}\rangle\rangle+(N+1)\,||\tilde{\Theta}\rangle\rangle\Big]\langle\langle\tilde{\Theta}||\bigg)_{r,r},\tag{S65}$$

where we use the definitions

$$||\Theta_{r,r'}\rangle\rangle=\frac{1}{\sqrt{g_{r,r'}}}\sum_{\substack{r_\alpha\in\{r,r+1\}\\ r_\beta\in\{r',r'+1\}}}\sum_{s_\alpha,s_\beta}||\alpha\beta\beta\alpha\rangle\rangle,\tag{S66}$$

$$||\bar{\Theta}_{r,r'}\rangle\rangle=\frac{1}{\sqrt{g_{r,r'}}}\sum_{\substack{r_\alpha\in\{r,r+1\}\\ r_\beta\in\{r',r'+1\}}}\sum_{s_\alpha,s_\beta}||\alpha\bar{\alpha}\beta\bar{\beta}\rangle\rangle,\tag{S67}$$

$$||\tilde{\Theta}_{r,r'}\rangle\rangle=\frac{1}{\sqrt{g_{r,r'}}}\sum_{\substack{r_\alpha\in\{r,r+1\}\\ r_\beta\in\{r',r'+1\}}}\sum_{s_\alpha,s_\beta}||\alpha\beta\alpha\beta\rangle\rangle,\tag{S68}$$

with $g_{r,r'} = N(N-\delta_{r,r'})$. Taking into account that $r$ and $r'$ have the same parity, these obey

$$\langle\langle i_{rr'}||\,i_{qq'}\rangle\rangle=\frac{16}{g_{r,r'}}\delta_{q,r}\delta_{q',r'}+\frac{4}{\sqrt{g_{rr'}g_{qq'}}}\delta_{q,r\pm1}\delta_{q',r'\pm1},\tag{S69}$$

$$\langle\langle i_{rr'}||\,j_{qq'}\rangle\rangle=\frac{\delta_{r,r'}}{6}\Big(2\delta_{q,r}\delta_{q',r'}+\delta_{q,r+1}\delta_{q',r'+1}+\delta_{q,r-1}\delta_{q',r'-1}\Big),\tag{S70}$$

for $i\neq j\in\{\Theta,\bar{\Theta},\tilde{\Theta}\}$. Notice that, while the evolution of $||\overline{O_0^2(t)}\rangle\rangle$ in $C_1(r,t)$ involved mixing within each pair of sites, $(r,r+1)$, through the projector $||\phi_r\rangle\rangle\langle\langle\phi_r||$, here the evolution through $v_{r,r'}$ entails mixing between each two pairs $(r,r+1)$ and $(r',r'+1)$. Thus, we expect the time evolution of $||\overline{O_0^2(t)}\rangle\rangle$ to be a two-dimensional process in space.

Finally, considering (S64), considering the average of (S60) becomes

$$||\overline{Q_0(t+1)}\rangle\rangle=\bigg(\sum_{R,R'=0}^{L/2-1}v_{2R,2R'}\bigg)\bigg(\sum_{R,R'=0}^{L/2-1}v_{2R-1,2R'-1}\bigg)||\overline{Q_0(t)}\rangle\rangle,\tag{S71}$$

with $v_{r,r'}$ given by (S65). Starting with $||\overline{Q_0(0)}\rangle\rangle$ given by (S58), we only need to apply this expression recursively to obtain $||\overline{Q_0(t)}\rangle\rangle$ at successive time steps. Next, we do precisely this and simplify (S71).

2. *Dynamics of $||\overline{Q_0(t)}\rangle\rangle$*

Let us start by defining

$$||\Theta\bar{\Theta}_{r,r'}\rangle\rangle = ||\Theta_{r,r'}\rangle\rangle - ||\bar{\Theta}_{r,r'}\rangle\rangle. \tag{S72}$$

Starting with $||\overline{Q_0(0)}\rangle\rangle$ given by (S58), applying the first odd layer of the circuit leads to

$$||\overline{Q_0(t=1/2)}\rangle\rangle = \left(\sum_{R,R'=0}^{L/2-1} v_{2R-1,2R'-1}\right)||Q_0(0)\rangle\rangle = v_{-1,-1}\,||Q_0(0)\rangle\rangle = \frac{1}{\sqrt{3}}||\Theta\bar{\Theta}_{-1,-1}\rangle\rangle, \tag{S73}$$

to which we can apply a second even layer to obtain

$$||\overline{Q_0(t=1)}\rangle\rangle = \left(\sum_{R,R'=0}^{L/2-1} v_{2R,2R'}\right)||\overline{Q_0(1/2)}\rangle\rangle = (v_{-2,-2} + v_{-2,0} + v_{0,-2} + v_{0,0})\,||\overline{Q_0(1/2)}\rangle\rangle$$

$$= \frac{1}{6\sqrt{3}}\Big(||\Theta\bar{\Theta}_{-2,-2}\rangle\rangle + ||\Theta\bar{\Theta}_{0,0}\rangle\rangle\Big) + \frac{1}{6}\Big(||\Theta\bar{\Theta}_{-2,0}\rangle\rangle + ||\Theta\bar{\Theta}_{0,-2}\rangle\rangle\Big). \tag{S74}$$

Also, we can use (S69) and (S70) to see that $||\Theta\bar{\Theta}_{r,r'}\rangle\rangle$ evolves under $v_{r,r'}$ according to

$$v_{r+a,r'+a}\,||\Theta\bar{\Theta}_{r,r'}\rangle\rangle = \left(\frac{1}{4} - \frac{\delta_{r,r'}}{12}\right)||\Theta\bar{\Theta}_{r+a,r'+a}\rangle\rangle, \tag{S75}$$

$$v_{r+a,r'-a}\,||\Theta\bar{\Theta}_{r,r'}\rangle\rangle = \left[\frac{1}{4} + \left(\frac{1}{2\sqrt{3}} - \frac{1}{4}\right)(\delta_{r,r'} + \delta_{r+a,r'-a})\right]||\Theta\bar{\Theta}_{r+a,r'-a}\rangle\rangle, \tag{S76}$$

where $a = \pm 1$. These relations control the mixing occurring between neighbouring pairs in successive time steps and ensure that the subspace $\{||\Theta\bar{\Theta}_{r,r'}\rangle\rangle\}$ is closed under time evolution. This, allied with (S73), allows the decomposition

$$||\overline{Q_0(t)}\rangle\rangle = \begin{cases} \sum_{R=0}^{L/2-1} \frac{1}{2}||\Theta\bar{\Theta}_{2R,2R'}\rangle\rangle\langle\langle\Theta\bar{\Theta}_{2R,2R'}||\overline{Q_0(t)}\rangle\rangle & \text{, for } t \text{ integer} \\ \sum_{R=0}^{L/2-1} \frac{1}{2}||\Theta\bar{\Theta}_{2R-1,2R'-1}\rangle\rangle\langle\langle\Theta\bar{\Theta}_{2R-1,2R'-1}||\overline{Q_0(t)}\rangle\rangle & \text{, for } t \text{ half-integer} \end{cases} \tag{S77}$$

where the factor $1/2$ arises from $\langle\langle\Theta\bar{\Theta}_{r,r'}||\Theta\bar{\Theta}_{q,q'}\rangle\rangle = 2\delta_{r,q}\delta_{r',q'}$ (where the indices are restricted to having the same parity). As $\{||\phi_r\rangle\rangle\}$ was the optimal basis on which to express $||\overline{O_0^2(t)}\rangle\rangle$, so it happens that $\{\frac{1}{\sqrt{2}}||\Theta\bar{\Theta}_{2R,2R'}\rangle\rangle \mid R, R' = 0, \ldots, L/2-1\}$ and $\{\frac{1}{\sqrt{2}}||\Theta\bar{\Theta}_{2R-1,2R'-1}\rangle\rangle \mid R, R' = 0, \ldots, L/2-1\}$ are the natural orthonormal basis on which to express $||\overline{Q_0(t)}\rangle\rangle$. Let us comment on the states $||\alpha\beta\mu\nu\rangle\rangle$ grouped under the same $||\Theta\bar{\Theta}_{r,r'}\rangle\rangle$. Due to PH symmetry, out of the three types of states surviving the average of Haar-distributed gates — $||\alpha\beta\beta\alpha\rangle\rangle$, $||\alpha\bar{\alpha}\beta\bar{\beta}\rangle\rangle$ and $||\alpha\beta\alpha\beta\rangle\rangle$ — only the first two survive; explicitly: $\langle\langle\alpha\beta\alpha\beta||Q_0(t)\rangle\rangle = -\langle\langle\bar{\alpha}\beta\bar{\alpha}\beta||Q_0(t)\rangle\rangle \Rightarrow \sum_{s_\alpha}\langle\langle\alpha\beta\alpha\beta||Q_0(t)\rangle\rangle = 0$. Furthermore, PH symmetry implies $\langle\langle\alpha\beta\beta\alpha|Q_0(t)\rangle\rangle = -\langle\langle\alpha\bar{\alpha}\beta\bar{\beta}|Q_0(t)\rangle\rangle = |\langle\alpha|O_0(t)|\beta\rangle|^2$ such that $||\Theta\bar{\Theta}_{r,r'}\rangle\rangle$ groups the states $||\Theta_{r,r'}\rangle\rangle$ and $-||\bar{\Theta}_{r,r'}\rangle\rangle$ which evolve equally. The structure appearing restricts the dynamics of $||Q_0(t)\rangle\rangle$ such that it is simpler than one might have guessed, with only $8L^2$ entries out of $(2L)^4$ being non-zero.

Having established the good basis to use, we can employ (S75) and (S76) to obtain, for $t \geq 1/2$,

$$\langle\langle\Theta\bar{\Theta}_{r,r'}||\overline{Q_0(t+1/2)}\rangle\rangle = \langle\langle\Theta\bar{\Theta}_{r,r'}||\left(v_{r-1,r'-1} + v_{r+1,r'+1} + v_{r-1,r'+1} + v_{r+1,r'-1}\right)||\overline{Q_0(t)}\rangle\rangle$$

$$= \text{tr}\left[m_{r,r'}\begin{pmatrix}\langle\langle\Theta\bar{\Theta}_{r-1,r'-1}|| & \langle\langle\Theta\bar{\Theta}_{r-1,r'+1}|| \\ \langle\langle\Theta\bar{\Theta}_{r+1,r'-1}|| & \langle\langle\Theta\bar{\Theta}_{r+1,r'+1}||\end{pmatrix}^T ||\overline{Q_0(t)}\rangle\rangle\right], \tag{S78}$$

where $m_{r,r'}$ is a coefficient matrix which assumes different values depending on $r$ and $r'$:

$$m_{r,r} = \begin{pmatrix} \frac{1}{6} & \frac{1}{2\sqrt{3}} \\ \frac{1}{2\sqrt{3}} & \frac{1}{6} \end{pmatrix} \quad , \quad m_{r,r+2} = m_{r,r-2}^T = \begin{pmatrix} \frac{1}{4} & \frac{1}{4} \\ \frac{1}{2\sqrt{3}} & \frac{1}{4} \end{pmatrix} \quad , \quad m_{r,r'} = \frac{1}{4}\begin{pmatrix} 1 & 1 \\ 1 & 1 \end{pmatrix}, \tag{S79}$$

with the last $m_{r,r'}$ being valid for $|r - r'| > 2$. In the bulk, i.e. for $|r - r'| > 2$, this evolution equation is an average over neighbouring sites, as depicted in Fig. S3. This is similar to what is shown to happen for $\overline{C_1(r,t)}$ in Fig. S2, where the process occurring is one instead of two-dimensional.

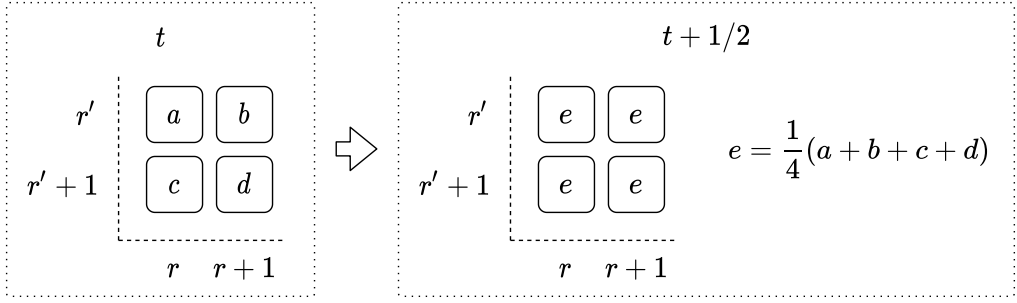

FIG. S3: Scheme of the evolution of the bulk $\langle\langle\Theta\bar\Theta_{r,r'}||\overline{Q_0(t)}\rangle\rangle|_{|r-r'|>2}$ (values in boxes) in one-half time step. The bulk evolution uses the third matrix $m_{r,r'}$ in (S79) and is effectively an average of neighbouring elements $\langle\langle\Theta\bar\Theta_{r,r'}||\overline{Q_0(t)}\rangle\rangle$ whose position indices belong to the pairs $(r, r+1)$ and $(r', r'+1)$. Depending on whether $t+1/2$ is integer or half-integer, these pairs are even or odd.

Applying (S78) twice gives

$$\langle\langle\Theta\bar\Theta_{r,r'}||\overline{Q_0(t+1)}\rangle\rangle = \text{tr}\left[M_{r,r'}\begin{pmatrix}\langle\langle\Theta\bar\Theta_{r-2,r'-2}|| & \langle\langle\Theta\bar\Theta_{r-2,r'}|| & \langle\langle\Theta\bar\Theta_{r-2,r'+2}|| \\ \langle\langle\Theta\bar\Theta_{r,r'-2}|| & \langle\langle\Theta\bar\Theta_{r,r'}|| & \langle\langle\Theta\bar\Theta_{r,r'+2}|| \\ \langle\langle\Theta\bar\Theta_{r+2,r'-2}|| & \langle\langle\Theta\bar\Theta_{r+2,r'}|| & \langle\langle\Theta\bar\Theta_{r+2,r'+2}||\end{pmatrix}^T ||\overline{Q_0(t)}\rangle\rangle\right] \quad (S80)$$

with $M_{r,r'}$ a coefficient matrix given by

$$M_{r,r} = \frac{1}{72}\begin{pmatrix}2 & 5\sqrt{3} & 3\sqrt{3} \\ 5\sqrt{3} & 16 & 5\sqrt{3} \\ 3\sqrt{3} & 5\sqrt{3} & 2\end{pmatrix} \quad,\quad M_{r,r+2} = M_{r,r-2}^T = \frac{1}{48}\begin{pmatrix}3 & 6 & 3 \\ \frac{10}{\sqrt{3}} & 13 & 6 \\ 4 & \frac{10}{\sqrt{3}} & 3\end{pmatrix}$$

$$M_{r,r+4} = M_{r,r-4}^T = \frac{1}{16}\begin{pmatrix}1 & 2 & 1 \\ 2 & 4 & 2 \\ \frac{2}{\sqrt{3}} & 2 & 1\end{pmatrix} \quad,\quad M_{r,r'} = \frac{1}{16}\begin{pmatrix}1 & 2 & 1 \\ 2 & 4 & 2 \\ 1 & 2 & 1\end{pmatrix}, \quad (S81)$$

where the last $M_{r,r'}$ is valid for $|r-r'|>4$. This last matrix establishes the evolution of the bulk of $\langle\langle\Theta\bar\Theta_{r,r'}||\overline{Q_0(t)}\rangle\rangle$ as a weighted average resulting from applying the scheme seen in Fig. S3 twice.

At last, starting with (S74) as our initial condition, $||\overline{Q_0(t)}\rangle\rangle$ is given by the recursive expression (S80) alongside (S77). This is equivalent to (S71) but more transparent than the latter. Now that we know the dynamics of $||\overline{Q_0(t)}\rangle\rangle$, we can obtain that of the OTOC.

### S.2. Prescription to obtain $\overline{C_2(r,t)}$

Now that we have reduced $||Q_0(t)\rangle\rangle$ to its core, we focus back on the $\overline{C_2(r,t)}$. At $t=0$, $||Q_0(0)\rangle\rangle$ is given by (S58) such that (S55) becomes

$$\overline{C_2(r,t=0)} = C_2(r,t=0) = \langle\langle Q_r||S||Q_0(0)\rangle\rangle = 2\delta_{r,0}. \quad (S82)$$

For $t\geq 1$, we can decompose $||\overline{Q_0(t)}\rangle\rangle$ using (S77) to obtain

$$\overline{C_2(r,t)} = \langle\langle Q_r||S||\overline{Q_0(t)}\rangle\rangle = \langle\langle Q_r||S\left(\sum_{R,R'=0}^{L/2-1}\frac{1}{2}||\Theta\bar\Theta_{2R,2R'}\rangle\rangle\langle\langle\Theta\bar\Theta_{2R,2R'}||\overline{Q_0(t)}\rangle\rangle\right)$$

$$= \frac{1}{2\sqrt{3}}\sum_{R=0}^{L/2-1}(\delta_{r,2R}+\delta_{r,2R+1})\langle\langle\Theta\bar\Theta_{2R,2R}||\overline{Q_0(t)}\rangle\rangle. \quad (S83)$$

For convenience, we can define

$$K_{r,r'}(t) = \frac{1}{2\sqrt{3}}\langle\langle\Theta\bar\Theta_{r,r'}||\overline{Q_0(t)}\rangle\rangle, \quad (S84)$$

whose diagonal gives the OTOC: $\overline{C_2(2R,t)} = \overline{C_2(2R+1,t)} = K_{2R,2R}(t)$. Thus, we can use (S74) to get the OTOC at $t=1$, which is

$$\overline{C_2(2R, t=1)} = \overline{C_2(2R+1, t=1)} = \frac{1}{18}\Big(\delta_{R,-1} + \delta_{R,0}\Big), \tag{S85}$$

and at subsequent time steps the dynamics of $\overline{C_2(r,t)}$ is determined by that of $\langle\langle\Theta\bar{\Theta}_{r,r'}||\overline{O_0^2(t)}\rangle\rangle \Leftrightarrow K_{r,r'}(t)$ we just saw to be given by (S80), i.e. for $t \geq 1$ we have

$$\overline{C_2(2R, t+1)} = \overline{C_2(2R+1, t+1)} = \text{diag}[K_{2R,2R'}(t+1)]$$

$$= \text{diag}\left\{\text{tr}\left[M_{2R,2R'}\begin{pmatrix}K_{2R-2,2R'-2} & K_{2R-2,2R'} & K_{2R-2,2R'+2}\\ K_{2R,2R'-2} & K_{2R,2R'} & K_{2R,2R'+2}\\ K_{2R+2,2R'-2} & K_{2R+2,2R'} & K_{2R+2,2R'+2}\end{pmatrix}^T(t)\right]\right\} \tag{S86}$$

Note that, although the OTOC is given by the diagonal of $K_{r,r'}(t)$, we must keep track of the whole matrix to learn its dynamics. Before proceeding to inspect the dynamics of $K_{r,r'}(t)$ in the continuum limit, we give a pictorial view in Fig. S4 of how $K_{r,r'}(t)$ (and thus, the OTOC) evolves in half-time steps.

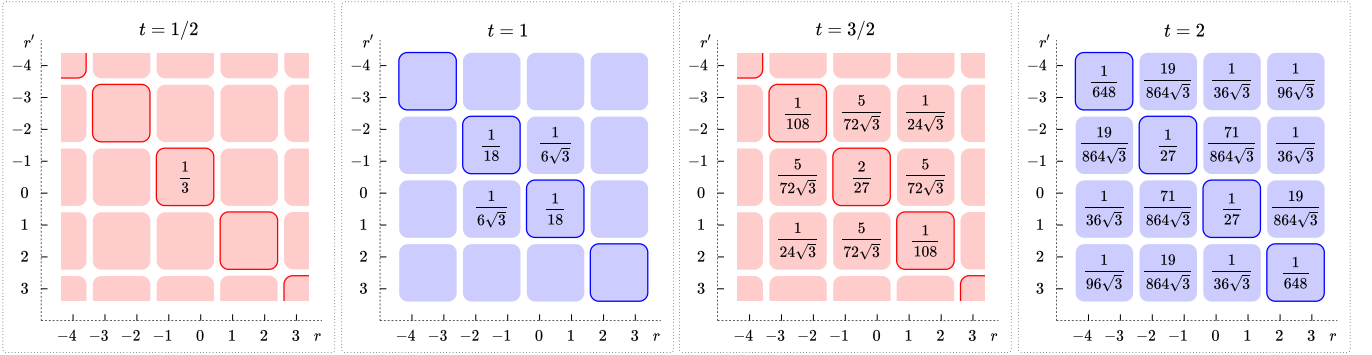

FIG. S4: The values in the boxes correspond to $K_{r,r'}(t)$ in the $r-r'$ plane, shown here for $t \in \{1/2, 1, 3/2, 2\}$ (empty boxes have zero value). Starting with $K_{r,r'}(t) = 2\delta_{r,r',0}$ as the initial condition, we evolve this by applying (S78) successively. The highlighted diagonal gives $\overline{C_2(r,t)}$. Note that the boxes group the pairs of sites $(r, r+1)$ and $(r', r'+1)$ enhancing the symmetry $K_{r,r'}(t) = K_{r+1,r'+1}(t) = K_{r+1,r'}(t) = K_{r,r'+1}(t)$. Depending on whether $t$ is integer or half-integer (blue/red), these pairs are even or odd.

### S.3. Continuum limit of $\overline{C_2(r,t)}$

We will see that the dynamics of $K_{r,r'}(t)$ is a diffusive process by showing that its evolution equation can be approximated by a two-dimensional diffusion equation in the continuum limit.

In the following, we obtain the continuum limit of (S78) which evolves $K_{r,r'}(t)$ in half time steps. Note that this equation (or, more specifically, the matrix $m_{r,r'}$) differs depending on whether $r' = r$, $r' = r \pm 2$ or $|r'-r| > 2$. However, since in the thermodynamic limit ($L \to \infty$) the regions with $r' = r$ and $r' = r \pm 2$ constitute a set of measure zero where $K_{r,r'}(t)$ does not diverge, they are not expected to influence the behaviour of the bulk. Thus, we start by obtaining a continuum diffusion equation for the bulk, i.e. for $K_{r,r'}(t)$ in $|r'-r| > 2$. Then, since we wish to obtain the OTOC, which is $K_{r,r}(t)$, we see that the diagonal differs from the bulk by a constant factor.

Consider the scaling form $a^2 K'(x = ra, x' = r'a, \tau = ta^2)$. For $a=1$, this coincides with the discrete $K_{r,r'}(t)$. If we let $a \to 0$, this approximates $K_{r,r'}(t)$ in the continuum limit, i.e. $\lim_{L \to \infty, t \to \infty} K_{r,r'}(t) = \lim_{a \to 0} a^2 K'(ra, r'a, ta^2)$. After making this identification in (S80) and considering $m_{r,r'}|_{|r-r'|>2}$ to be the isotropic bulk coefficient matrix given in (S79), we Taylor expand $K'_{\text{bulk}}(x, x', \tau)$ (i.e. the quantity $K'(x, x', \tau)$ evolving isotropically under (S80) with $m_{r,r'}|_{|r-r'|>2}$) up to $\mathcal{O}(a^2)$ such that (S78) becomes

$$\partial_\tau K'_{\text{bulk}}(x, x', \tau) = D_2(\partial_x^2 + \partial_{x'}^2) K'_{\text{bulk}}(x, x', \tau), \tag{S87}$$

i.e. the isotropic diffusion equation in 2D, with diffusion constant $D_2 = 1$.

We return to the question of the diagonal, $K_{r,r}(t)$. Namely, we will account for the anisotropy present in the evolution of the diagonal which will impact the way the initial condition, which is precisely in the diagonal, will be distributed along different directions, something which was disregarded to obtain (S87). To begin with, we distinguish between $r = r'$ and $r \neq r'$ and approximate $K_{r,r}(t)$ by $a^2 K'(x, x, \tau)$ for $r' = r$ and by $K'_{\text{bulk}}(x, x', \tau)$ for $r' \neq r$. Making this identification in (S78) gives $K'_{\text{bulk}}(x - a, x + a, \tau + 1/2) = 1/4[K'_{\text{bulk}}(x - 2a, x, \tau) + K'_{\text{bulk}}(x - 2a, x + 2a, \tau) + K'_{\text{bulk}}(x, x + 2a, \tau)] + 1/(2\sqrt{3}) K'(x, x, \tau)$ which, after Taylor expanding $K'_{\text{bulk}}$ up to the lowest order, $\mathcal{O}(a^0)$, yields

$$K'(x, x, \tau) \approx c K'_{\text{bulk}}(x, x, \tau), \tag{S88}$$

with $c = \sqrt{3}/2$. Note that we approximated $K'(x, x \pm 2a, \tau)$ by the bulk $K'_{\text{bulk}}$: similarly to what is done to obtain (S88), we can Taylor expand $K'_{\text{bulk}}$ up to $\mathcal{O}(a^0)$ in the evolution equation of $K'_{\text{bulk}}(x - 2a, x + 2a, \tau + 1/2)$ to obtain $K'(x, x \pm 2a, \tau) \sim K'_{\text{bulk}}(x, x' \pm 2a, \tau)$.

Considering (S88) in the evolution equation of the diagonal (S78) leads to the weighted average $K'_{\text{bulk}}(x, x, \tau + 1/2) =$

$$\text{tr}\left[m'_{r,r'} \begin{pmatrix} K'_{\text{bulk}}(x - 2, x - 2, \tau) & K'_{\text{bulk}}(x - 2, x + 2, \tau) \\ K'_{\text{bulk}}(x + 2, x - 2, \tau) & K'_{\text{bulk}}(x + 2, x + 2, \tau) \end{pmatrix}^T \right] \quad \text{with} \quad m'_{r,r} = \frac{1}{6}\begin{pmatrix} 1 & 2 \\ 2 & 1 \end{pmatrix}. \tag{S89}$$

This allows us to directly compare the diagonal with the bulk. The anisotropy in the new coefficient matrix $m'_{r,r}$ leads to diffusion with weighting factor $f = 1/2$ (the anti-diagonal elements are twice the diagonal ones). Thus, to obtain $K'(x, x, \tau)$, we take the diagonal of $K'_{\text{bulk}}(x, x', \tau)$ given as the solution of (S87) and multiply it by $c$ — which accounts for the different scaling of the diagonal — and by $f$ — which accounts for the anisotropy present.

Finally, we can obtain the solution of (S87) and related it to the OTOC. The initial condition (S82) translates to $K'(x, x, \tau = 0) \approx c K'_{\text{bulk}}(x, x, \tau = 0) = A\delta(x)\delta(x')$, with $A = 2$ and where we used (S88), such that the solution to (S87) is

$$c K'_{\text{bulk}}(x, x', \tau) = \frac{A}{4\pi\tau D_2} \exp\left(-\frac{x^2 + x'^2}{4 D_2 \tau}\right). \tag{S90}$$

Finally, the OTOC is approximated by (S90) times the corrective factor $f$

$$\overline{C_2(2R, t)} = \overline{C_2(2R+1, t)} = K_{2R,2R}(t) \simeq fca^2 K'_{\text{bulk}}(2Ra, 2Ra, \tau) = \frac{A(t)}{\sqrt{2\pi}\sigma(t)} \exp\left(-\frac{(2R)^2}{2\sigma(t)^2}\right), \tag{S91}$$

i.e. in the continuum limit, the OTOC is described by a Gaussian with broadening width $\sigma(t) = \sqrt{t}$ and varying normalization $A(t) = fA/(2\sqrt{2\pi t})$, with $A = 2$ and $f = 1/2$. We conclude that $\overline{C_2(r, t)}$ is the diagonal of a quantity which diffuses in 2D, analogously to $\overline{C_1(r, t)}$ which instead diffused in 1D.

### IX.  NC-T: AVERAGED TOC, $\overline{C_1(r, t)}$

We now turn to the second instance of free fermion evolution, NC-T, which, like the NC-ST, case does not conserve particle number and, unlike it, it displays randomness only in time. In this circuit, each layer $U_{\text{even}}$ and $U_{\text{odd}}$ is built acting with the same randomly chosen gate $u$ on all pairs of sites such that the system is invariant under space translation by multiples of two sites, as shown drawn in Fig. S1 (b). Next, we derive the behaviour of $\overline{C_1(r, t)}$ in this case by pointing out the differences from the NC-ST case, which reside in the average of unitaries. We saw in Sec. VII S.1 that the average of unitaries $\overline{u^\dagger_{r,r+1} \otimes u^T_{r',r'+1}}$ is non-zero if $u_{r,r+1} = u_{r',r'+1}$. While in the NC-ST case this demanded $r' = r$, the spatial homogeneity of NC-T softens this condition to $r' = r$ modulo 2. As a consequence, the operator $\sum_r ||\phi_r\rangle\rangle \langle\langle\phi_r||$ which evolves $||\overline{O_0^2(t)}\rangle\rangle$ in the NC-ST is generalized to $\sum_{rr'} ||\phi_{rr'}\rangle\rangle \langle\langle\phi_{rr'}||$ with

$$||\phi_r\rangle\rangle \longrightarrow ||\phi_{rr'}\rangle\rangle = \frac{1}{2} \sum_{b \in \{0,1\}} \sum_s ||r + b, s; r' + b, s\rangle\rangle. \tag{S92}$$

Considering the new evolution operator, we can start from $||\overline{O_0^2(0)}\rangle\rangle = \sum_s ||0, s; 0, s\rangle\rangle$ to obtain

$$||\overline{Q_0(1/2)}\rangle\rangle = \left(\sum_{R,R'=0}^{L/2-1} ||\phi_{2R,2R'}\rangle\rangle \langle\langle\phi_{2R,2R'}||\right) ||\overline{Q_0(0)}\rangle\rangle = ||\phi_{-1-1}\rangle\rangle = ||\phi_{-1}\rangle\rangle. \tag{S93}$$

The interaction between neighbouring pairs in successive layers analogous to (S46) reduces to

$$||\phi_{r,r'}\rangle\rangle \langle\langle\phi_{r,r'}||\phi_{r+a,r'+a}\rangle\rangle = \frac{1}{2}||\phi_{r,r'}\rangle\rangle, \tag{S94}$$

with $a = \pm 1$. From this, we get an evolution equation analogous to (S48)

$$\langle\langle\phi_{r,r'}||\overline{O_0^2(t+1/2)}\rangle\rangle = \langle\langle\phi_{r,r'}||\Big(||\phi_{r-1,r'-1}\rangle\rangle\langle\langle\phi_{r-1,r'-1}|| + ||\phi_{r+1,r'+1}\rangle\rangle\langle\langle\phi_{r+1,r'+1}||\Big)||\overline{O_0^2(t)}\rangle\rangle$$
$$= \frac{1}{2}\Big(\langle\langle\phi_{r-1,r'-1}|| + \langle\langle\phi_{r+1,r'+1}||\Big)||\overline{O_0^2(t)}\rangle\rangle. \tag{S95}$$

This, alongside the initial condition (S93), leads only terms of the type $||\phi_{r,r}\rangle\rangle = ||\phi_r\rangle\rangle$ to arise. Thus, the above expression is effectively reduced to (S48) which, allied with the initial condition $||\overline{Q_0(1/2)}\rangle\rangle = ||\phi_{-1}\rangle\rangle$ being the same we had for NC-ST, leads $\overline{C_1(r,t)}$ in the NC-T case to be exactly equal to the one obtained for NC-ST. Although the spatial homogeneity of the new circuit threatens the TOC dynamics to be described by a two-dimensional process, the initial condition fixes it to the one-dimensional process already known for NC-ST.

## X. NC-T: AVERAGED OTOC, $\overline{C_2(r,t)}$

We continue to consider the NC-T case, focusing now on $\overline{C_2(r,t)}$. As happens with $\overline{C_1(r,t)}$, the difference from the NC-ST case lies in the average of unitaries.

We saw in Sec. VIII S.1 that, for the average $\overline{u_{r_1,r_1+1}^\dagger \otimes u_{r_2,r_2+1}^\dagger \otimes u_{r_1',r_1'+1}^T \otimes u_{r_2',r_2'+1}^T}$ to be non-zero, either two or four unitaries must be equal, which, in the NC-ST case, translates to $r_1 = r_2 \wedge r_1' = r_2'$ or $r_1 = r_1' \wedge r_2 = r_2'$ or $r_1 = r_2' \wedge r_2 = r_1'$ or $r_1 = r_2 = r_1' = r_2'$. Similarly to what we just saw for $\overline{C_1(r,t)}$, these restrictions are softened in the NC-T case: since all the gates composing one layer are the same the four unitaries can refer to different pairs such that $v_{r,r'}$ is generalized to $v_{r_1,r_2,r_1',r_2'}$ by transforming (S66), (S67) and (S68) according to

$$||\Theta_{r,r'}\rangle\rangle \to ||\Theta_{r_1,r_2,r_1',r_2'}\rangle\rangle = \frac{1}{\sqrt{g_{r_1,r_2}}} \sum_{\substack{b_1,b_2\in\{0,1\} \\ s_1,s_2}} ||r_1+b_1,s_1;r_2+b_2,s_2;r_2'+b_2,s_2;r_1'+b_1,s_1\rangle\rangle, \tag{S96}$$

$$||\bar{\Theta}_{r,r'}\rangle\rangle \to ||\bar{\Theta}_{r_1,r_2,r_1',r_2'}\rangle\rangle = \frac{1}{\sqrt{g_{r_1,r_2}}} \sum_{\substack{b_1,b_2\in\{0,1\} \\ s_1,s_2}} ||r_1+b_1,s_1;r_1'+b_1,\bar{s}_1;r_2+b_2,s_2;r_2'+b_2,\bar{s}_2\rangle\rangle, \tag{S97}$$

$$||\tilde{\Theta}_{r,r'}\rangle\rangle \to ||\tilde{\Theta}_{r_1,r_2,r_1',r_2'}\rangle\rangle = \frac{1}{\sqrt{g_{r_1,r_2}}} \sum_{\substack{b_1,b_2\in\{0,1\} \\ s_1,s_2}} ||r_1+b_1,s_1;r_2+b_2,s_2;r_1'+b_1,s_1;r_2'+b_2,s_2\rangle\rangle. \tag{S98}$$

These obey relations equivalent to (S69) and (S70),

$$\langle\langle i_{r_1,r_2,r_1',r_2'}||i_{q_1,q_2,q_1',q_2'}\rangle\rangle = \frac{16}{g_{r_1,r_2}}\delta_{q_1 r_1}\delta_{q_2 r_2}\delta_{q_1' r_1'}\delta_{q_2' r_2'} + \frac{4}{\sqrt{g_{r_1 r_2}g_{q_1 q_2}}}\delta_{q_1,r_1+a}\delta_{q_2,r_2+b}\delta_{q_1',r_1'+a}\delta_{q_2',r_2'+b}, \tag{S99}$$

$$\langle\langle i_{r_1,r_2,r_1',r_2'}||j_{q_1,q_2,q_1',q_2'}\rangle\rangle = \frac{\delta_{r_1 r_2}}{6}\Big(2\delta_{q_1 r_1}\delta_{q_2 r_2}\delta_{q_1' r_1'}\delta_{q_2' r_2'} + \delta_{q_1,r_1+1}\delta_{q_2,r_2+1}\delta_{q_1',r_1'+1}\delta_{q_2',r_2'+1}$$
$$+ \delta_{q_1,r_1-1}\delta_{q_2,r_2-1}\delta_{q_1',r_1'-1}\delta_{q_2',r_2'-1}\Big), \tag{S100}$$

with $a, b = \pm 1$, which can be used to obtain the interaction between neighbouring pairs in successive layers

$$v_{r_1+a,r_2+a,r_1'+a,r_2'+a}||\Theta\bar{\Theta}_{r_1,r_2,r_1',r_2'}\rangle\rangle = \left(\frac{1}{4} - \frac{\delta_{r,r'}}{12}\right)||\Theta\bar{\Theta}_{r_1+a,r_2+a,r_1'+a,r_2'+a}\rangle\rangle, \tag{S101}$$

$$v_{r_1+a,r_2-a,r_1'+a,r_2'-a}||\Theta\bar{\Theta}_{r_1,r_2,r_1',r_2'}\rangle\rangle = \left[\frac{1}{4} + \left(\frac{1}{2\sqrt{3}} - \frac{1}{4}\right)(\delta_{r,r'} + \delta_{r+a,r'-a})\right]||\Theta\bar{\Theta}_{r_1+a,r_2-a,r_1'+a,r_2'-a}\rangle\rangle, \tag{S102}$$

Applying this to $||Q_0(0)\rangle\rangle$, given by (S58), results in

$$||\overline{Q_0(t=1/2)}\rangle\rangle = \left(\sum_{R_1,R_2,R_1',R_2'=0}^{L/2-1} v_{2R_1-1,2R_2-1,2R_1'-1,2R_2'-1}\right)||Q_0(0)\rangle\rangle$$
$$= v_{-1,-1,-1,-1}||Q_0(0)\rangle\rangle = \frac{1}{\sqrt{3}}||\Theta\bar{\Theta}_{-1,-1,-1,-1}\rangle\rangle = \frac{1}{\sqrt{3}}||\Theta\bar{\Theta}_{-1,-1}\rangle\rangle, \tag{S103}$$

the same initial condition we have for the NC-ST case. Since in (S101) and (S102) $r_1$ and $r_1'$ as well as $r_2$ and $r_2'$ are restricted to varying by the same amount, the subspace $\{||\Theta\bar{\Theta}_{r_1,r_2,r_1,r_2}\rangle\rangle\}$ is closed under time evolution. Thus, starting with (S103) as our initial condition, the relation $||\Theta\bar{\Theta}_{r_1,r_2,r_1,r_2}\rangle\rangle = ||\Theta\bar{\Theta}_{r_1,r_2}\rangle\rangle$ effectively reduces the evolution relations (S101) and (S102) to the ones we had for NC-ST, (S75) and (S76). It follows that $\overline{C_2(r,t)}$ for the NC-T case is equal to the one obtained before for the NC-ST case. Equivalently to what happens for $\overline{C_1(r,t)}$, although the spatial homogeneity of the new circuit threatens the OTOC dynamics to be described by a four-dimensional process, initial conditions fix it to the two-dimensional process already known for NC-ST.

## XI. C-ST: AVERAGED TOC, $\overline{C_1(r,t)}$

Lastly, we focus on the third instance of free fermion evolution, C-ST, a restriction of the NC-ST case to a particle conserving process admitting spatio-temporal noise. The difference from NC-ST lies in the new two-site unitaries conserving particle number, i.e. we have no anomalous terms $a_i a_j$ or $a_i^\dagger a_j^\dagger$ such that the $ph$ and $hp$ sectors of $u$ are null (by $ph$ sector we mean those $|r,s\rangle\langle r,s|u|r',s'\rangle\langle r',s'|$ with $s = p$ and $s' = h$). The new unitaries are diagonal by blocks: $u = \text{diag}(w,w)$, with $w$ $2 \times 2$ random Haar-distributed unitary matrices. We start by computing $\overline{C_1(r,t)}$ and then $\overline{C_2(r,t)}$.

Knowing that $\overline{w^*_{\alpha\beta} w_{\mu\nu}} = \frac{1}{N}\delta_{\alpha\mu}\delta_{\beta\nu}$ according to (S31) in Sec. V, the average equivalent to (S40) becomes

$$\overline{\langle\alpha_1|u|\beta_1\rangle^* \langle\alpha_2|u|\beta_2\rangle} = \overline{\langle\alpha_1|w|\beta_1\rangle^* \langle\alpha_2|w|\beta_2\rangle}\delta_{s_{\alpha_1}s_{\beta_1}}\delta_{s_{\alpha_2}s_{\beta_2}} = \frac{1}{N}\delta_{\alpha_1\alpha_2}\delta_{\beta_1\beta_2}\delta_{s_{\alpha_1}s_{\beta_1}}\delta_{s_{\alpha_2}s_{\beta_2}}, \tag{S104}$$

where $N = \text{rank}(w) = 2$, with the difference from the NC-ST being a restriction to the non-anomalous sectors $pp$ and $hh$. This results in the evolution operator (S41) becoming

$$||\phi_r\rangle\rangle\langle\langle\phi_r|| \longrightarrow \Phi_r = \frac{1}{2}\sum_{\substack{x\in\{r,r+1\}\\x'\in\{r',r'+1\}}}\sum_s ||x,s;x,s\rangle\rangle\langle\langle x',s;x',s||. \tag{S105}$$

The initial condition $||\overline{O_0^2(0)}\rangle\rangle = \sum_s ||0,s;0,s\rangle\rangle$ evolves under this as

$$||\overline{Q_0(1/2)}\rangle\rangle = \left(\sum_{R=0}^{L/2-1}\Phi_{2R-1}\right)||\overline{Q_0(0)}\rangle\rangle = \Phi_{-1}||\overline{Q_0(0)}\rangle\rangle = ||\phi_{-1}\rangle\rangle, \tag{S106}$$

i.e. $||\overline{Q_0(1/2)}\rangle\rangle$ is equal to the one we had for both NC-ST and NC-T. Also, the interaction between neighbouring pairs in successive layers reduces to

$$\Phi_{r\pm 1}||\phi_r\rangle\rangle = \frac{1}{2}||\phi_{r\pm 1}\rangle\rangle, \tag{S107}$$

also analogous to $||\phi_r\rangle\rangle\langle\langle\phi_{r'}|\phi_{r\pm 1}\rangle\rangle = 1/2$ valid for the NC-ST case. It follows that $\overline{C_1(r,t)}$ is the same we obtained for the other two cases, with the restriction in the particle-hole state in (S105), when comparing to $||\phi_r\rangle\rangle\langle\langle\phi_r||$, being compensated by the different prefactor.

## XII. C-ST: AVERAGED OTOC, $\overline{C_2(r,t)}$

While the TOC and OTOC obtained previously for the NC-T and C-ST cases match the results obtained for NC-ST, the OTOC for C-ST does not. Next, we obtain precisely $\overline{C_2(r,t)}$ for the C-ST case, pointing out the differences from its non-conserving counterpart.

### S.1. Time evolution of $||\overline{Q_0(t)}\rangle\rangle$

As done for the NC-ST case, we start by computing $||\overline{Q_0(t)}\rangle\rangle$.

### 1. Average over disorder realizations

Particle conserving free fermion gates vary from their non-conserving counterparts. Being built as mentioned at the beginning of Sec. XI, their moments can be obtained directly from those of $2 \times 2$ unitary matrices, $w$. Thus, their second moments $\overline{\langle\beta_1|\,u\,|\alpha_1\rangle^*\,\langle\beta_2|\,u\,|\alpha_2\rangle}$ are given by (S104) while its fourth moment is given by

$$\overline{\langle\beta_1|\,u\,|\alpha_1\rangle^*\,\langle\beta_2|\,u\,|\alpha_2\rangle^*\,\langle\beta_3|\,u\,|\alpha_3\rangle\,\langle\beta_4|\,u\,|\alpha_4\rangle}$$
$$= \overline{\langle\beta_1|\,w\,|\alpha_1\rangle^*\,\langle\beta_2|\,w\,|\alpha_2\rangle^*\,\langle\beta_3|\,w\,|\alpha_3\rangle\,\langle\beta_4|\,w\,|\alpha_4\rangle}\delta_{s_{\alpha_1}s_{\beta_1}}\delta_{s_{\alpha_2}s_{\beta_2}}\delta_{s_{\alpha_3}s_{\beta_3}}\delta_{s_{\alpha_4}s_{\beta_4}}, \quad (S108)$$

with the average of unitary matrices being (see Corollary 2.4 in [4])

$$\overline{\langle\beta_1|\,w\,|\alpha_1\rangle^*\,\langle\beta_2|\,w\,|\alpha_2\rangle^*\,\langle\beta_3|\,w\,|\alpha_3\rangle\,\langle\beta_4|\,w\,|\alpha_4\rangle}$$
$$= \frac{1}{N(N^2-1)}\Big[\delta_{\beta_1\beta_4}\delta_{\beta_2\beta_3}(N\delta_{\alpha_1\alpha_4}\delta_{\alpha_2\alpha_3} - \delta_{\alpha_1\alpha_3}\delta_{\alpha_2\alpha_4}) + \delta_{\beta_1\beta_3}\delta_{\beta_2\beta_4}(N\delta_{\alpha_1\alpha_3}\delta_{\alpha_2\alpha_4} - \delta_{\alpha_1\alpha_4}\delta_{\alpha_2\alpha_3})\Big]. \quad (S109)$$

with $N = \text{rank}(w) = 2$. With these changes, $v_{r,r'}$ given previously by (S65) becomes

$$v_{r,r'} \to v_{r,r'}^C = \frac{1-\delta_{rr'}}{N^2}\Big[||\theta\rangle\rangle\,\langle\langle\theta|| \otimes (\Pi + \pi) + ||\tilde{\theta}\rangle\rangle\,\langle\langle\tilde{\theta}|| \otimes (\Pi + \tilde{\pi})\Big]_{r,r'}$$
$$+ \frac{\delta_{rr'}}{N(N^2-1)}\Big[N\,||\theta\rangle\rangle\,\langle\langle\theta|| \otimes (\Pi + \pi) - ||\theta\rangle\rangle\,\langle\langle\tilde{\theta}|| \otimes \Pi - ||\tilde{\theta}\rangle\rangle\,\langle\langle\theta|| \otimes \Pi + N\,||\tilde{\theta}\rangle\rangle\,\langle\langle\tilde{\theta}|| \otimes (\Pi + \tilde{\pi})\Big]_{r,r}, \quad (S110)$$

where we used the definitions

$$||\theta_{r,r'}\rangle\rangle = \sum_{x\in\{r,r+1\}}\sum_{x'\in\{r',r'+1\}}||x,x',x',x\rangle\rangle, \quad (S111)$$

$$||\tilde{\theta}_{r,r'}\rangle\rangle = \sum_{x\in\{r,r+1\}}\sum_{x'\in\{r',r'+1\}}||x,x',x,x'\rangle\rangle, \quad (S112)$$

which respect

$$\langle\langle i_{rr'}||\,i_{qq'}\rangle\rangle = 4\delta_{q,r}\delta_{q',r'} + \delta_{q,r\pm 1}\delta_{q',r'\pm 1}, \quad (S113)$$

$$\langle\langle i_{rr'}||\,j_{qq'}\rangle\rangle = \delta_{r,r'}\Big(2\delta_{q,r}\delta_{q',r'} + \delta_{q,r+1}\delta_{q',r'+1} + \delta_{q,r-1}\delta_{q',r'-1}\Big), \quad (S114)$$

with $i \neq j \in \{\theta, \bar{\theta}, \tilde{\theta}\}$. We also used

$$\Pi = ||pppp\rangle\rangle\,\langle\langle pppp|| + ||hhhh\rangle\rangle\,\langle\langle hhhh|| \quad , \quad ||\Pi\rangle\rangle = ||pppp\rangle\rangle + ||hhhh\rangle\rangle \quad (S115)$$
$$\pi = ||phhp\rangle\rangle\,\langle\langle phhp|| + ||hpph\rangle\rangle\,\langle\langle hpph|| \quad , \quad ||\pi\rangle\rangle = ||phhp\rangle\rangle + ||hpph\rangle\rangle \quad (S116)$$
$$\tilde{\pi} = ||phph\rangle\rangle\,\langle\langle phph|| + ||hphp\rangle\rangle\,\langle\langle hphp|| \quad , \quad ||\tilde{\pi}\rangle\rangle = ||phph\rangle\rangle + ||hphp\rangle\rangle \quad (S117)$$

for which $PP' = \delta_{PP'}$ and $P\,||P\rangle\rangle = ||P\rangle\rangle$ holds, with $P$ and $P' \in \{\pi, \bar{\pi}, \tilde{\pi}\}$.

### 2. Dynamics of $||\overline{Q_0(t)}\rangle\rangle$

Defining

$$||\Theta\tilde{\Theta}_{r,r'}\rangle\rangle = \Big(||\theta_{r,r'}\rangle\rangle + ||\tilde{\theta}_{r,r'}\rangle\rangle\Big) \otimes ||\Pi\rangle\rangle, \quad (S118)$$

we obtain from initial conditions

$$||\overline{Q_0(t=1/2)}\rangle\rangle = v_{-1,-1}^C\,||Q_0(0)\rangle\rangle = \frac{1}{6}\,||\Theta\tilde{\Theta}_{-1,-1}\rangle\rangle - \frac{1}{3}\,||\tilde{\theta}_{-1,-1}\rangle\rangle \otimes ||\tilde{\pi}\rangle\rangle, \quad (S119)$$

to which we can apply a second even layer of gates to obtain

$$||\overline{Q_0(t=1)}\rangle\rangle = \left(\sum_{R,R'=0}^{L/2-1} v_{2R,2R'}^C\right) ||Q_0(1/2)\rangle\rangle = (v_{-2,-2}^C + v_{-2,0}^C + v_{0,-2}^C + v_{0,0}^C) ||Q_0(1/2)\rangle\rangle$$
$$= \frac{1}{18}\left(||\Theta\tilde{\Theta}_{-2,-2}\rangle\rangle + ||\Theta\tilde{\Theta}_{0,0}\rangle\rangle\right) + \frac{1}{24}\left(||\Theta\tilde{\Theta}_{-2,0}\rangle\rangle + ||\Theta\tilde{\Theta}_{0,-2}\rangle\rangle\right)$$
$$- \frac{1}{9}\left(||\tilde{\theta}_{-2,-2}\rangle\rangle + ||\tilde{\theta}_{0,0}\rangle\rangle\right) \otimes ||\tilde{\pi}\rangle\rangle - \frac{1}{12}\left(||\tilde{\theta}_{-2,0}\rangle\rangle + ||\tilde{\theta}_{0,-2}\rangle\rangle\right) \otimes ||\tilde{\pi}\rangle\rangle. \quad (S120)$$

One can check, making use of (S113) and (S114), that $||\Theta\tilde{\Theta}_{r,r'}\rangle\rangle$ and $||\tilde{\theta}_{r,r'}\rangle\rangle \otimes ||\tilde{\pi}\rangle\rangle$ evolve under $v_{r,r'}^C$ as

$$v_{r+a,r'+a}^C ||\Theta\tilde{\Theta}_{r,r'}\rangle\rangle = \left(\frac{1}{4} + \frac{\delta_{r,r'}}{12}\right) ||\Theta\tilde{\Theta}_{r+a,r'+a}\rangle\rangle \quad (S121)$$

$$v_{r+a,r'-a}^C ||\Theta\tilde{\Theta}_{r,r'}\rangle\rangle = \left(\frac{1}{4} - \frac{\delta_{r,r'}}{12}\right) ||\Theta\tilde{\Theta}_{r+a,r'-a}\rangle\rangle \quad (S122)$$

$$v_{r\pm a,r'\pm a}^C ||\tilde{\theta}_{r,r'}\rangle\rangle \otimes ||\tilde{\pi}\rangle\rangle = \left(\frac{1}{4} + \frac{\delta_{r,r'}}{12}\right) ||\tilde{\theta}_{r\pm a,r'\pm a}\rangle\rangle \otimes ||\tilde{\pi}\rangle\rangle \quad (S123)$$

with $a = \pm 1$. These relations control the mixing occurring between neighbouring pairs in successive time steps and imply that both $\{||\Theta\tilde{\Theta}_{r,r'}\rangle\rangle\}$ and $\{||\tilde{\theta}_{r,r'}\rangle\rangle \otimes ||\tilde{\pi}\rangle\rangle\}$ are closed under time evolution, besides being orthogonal to one another (since $\langle\langle\Pi||\tilde{\pi}\rangle\rangle = 0$). Thus, we can decompose

$$||\overline{Q_0(t)}\rangle\rangle = \sum_{R=0}^{L/2-1} \left(\frac{||\Theta\tilde{\Theta}_{2R,2R'}\rangle\rangle \langle\langle\Theta\tilde{\Theta}_{2R,2R'}||}{8(2+\delta_{2R,2R'})} + \frac{||\tilde{\theta}_{2R,2R'}\rangle\rangle \langle\langle\tilde{\theta}_{2R,2R'}|| \otimes ||\tilde{\pi}\rangle\rangle \langle\langle\tilde{\pi}||}{8}\right) ||\overline{Q_0(t)}\rangle\rangle, \quad (S124)$$

valid for $t$ integer, becoming valid for $t$ half-integer after transforming $2R \to 2R-1$ and $2R' \to 2R'-1$. The prefactors come from $\langle\langle\Theta\tilde{\Theta}_{r,r'}||\Theta\tilde{\Theta}_{q,q'}\rangle\rangle = 8(2+\delta_{r,r'})\delta_{r,q}\delta_{r',q'}$ and $\langle\langle\tilde{\theta}_{r,r'}||\tilde{\theta}_{q,q'}\rangle\rangle \langle\langle\tilde{\pi}||\tilde{\pi}\rangle\rangle = 8\delta_{r,q}\delta_{r',q'}$ (where the indices are restricted to having the same parity). Ahead, we will see that only states of the type $||\Theta\tilde{\Theta}_{r,r'}\rangle\rangle$ contribute to the OTOC and thus we can ignore $\{||\tilde{\theta}_{r,r'}\rangle\rangle \otimes ||\tilde{\pi}\rangle\rangle\}$. Although $||\Theta\tilde{\Theta}_{r,r'}\rangle\rangle$ are analogous to the states $||\Theta\bar{\Theta}_{r,r'}\rangle\rangle$ appearing in the NC-ST case, the two are inherently different: while $||\Theta\tilde{\Theta}_{r,r'}\rangle\rangle$ includes, besides $||\alpha\beta\beta\alpha\rangle\rangle$, states of the type $||\alpha\beta\alpha\beta\rangle\rangle$ which do not appear in NC-ST due to PH symmetry, it leaves out states of the kind $||\alpha\bar{\alpha}\beta\bar{\beta}\rangle\rangle$, which do not appear when averaging over unitary gates.

Having $\{\frac{1}{8(2+\delta_{r,r'})} ||\Theta\tilde{\Theta}_{r,r'}\rangle\rangle\}$ as the natural basis to use, we can employ (S121) and (S122) to obtain, for $t \geq 1/2$, an expression equivalent to (S78) with the changes $\langle\langle\Theta\bar{\Theta}_{r,r'}|| \to \langle\langle\Theta\tilde{\Theta}_{r,r'}||$ and $m_{r,r'} \to m_{r,r'}^C$, with $m_{r,r'}^C$ coefficient matrices given by

$$m_{r,r}^C = \begin{pmatrix} \frac{1}{3} & \frac{1}{4} \\ \frac{1}{4} & \frac{1}{3} \end{pmatrix} \quad , \quad m_{r,r+2}^C = (m_{r,r-2}^C)^T = \begin{pmatrix} \frac{1}{4} & \frac{1}{4} \\ \frac{1}{6} & \frac{1}{4} \end{pmatrix} \quad , \quad m_{r,r'}^C = \frac{1}{4}\begin{pmatrix} 1 & 1 \\ 1 & 1 \end{pmatrix}, \quad (S125)$$

with the last $m_{r,r'}^C$ being valid for $|r-r'| > 2$. Applying this twice gives an expression equivalent to (S80) with $\langle\langle\Theta\bar{\Theta}_{r,r'}|| \to \langle\langle\Theta\tilde{\Theta}_{r,r'}||$ and $M_{r,r'} \to M_{r,r'}^C$, with the new coefficient matrices $M_{r,r'}^C$ being

$$M_{r,r}^C = \frac{1}{144}\begin{pmatrix} 16 & 21 & 9 \\ 21 & 44 & 21 \\ 9 & 21 & 16 \end{pmatrix} \quad , \quad M_{r,r+2}^C = (M_{r,r-2}^C)^T = \frac{1}{32}\begin{pmatrix} 1 & 2 & 1 \\ \frac{14}{9} & \frac{11}{3} & 2 \\ \frac{2}{3} & \frac{14}{9} & 1 \end{pmatrix}$$
$$M_{r,r+4}^C = (M_{r,r-4}^C)^T = \frac{1}{16}\begin{pmatrix} 1 & 2 & 1 \\ 2 & 4 & 2 \\ \frac{2}{3} & 2 & 1 \end{pmatrix} \quad , \quad M_{r,r'}^C = \frac{1}{16}\begin{pmatrix} 1 & 2 & 1 \\ 2 & 4 & 2 \\ 1 & 2 & 1 \end{pmatrix}, \quad (S126)$$

where the last $M_{r,r'}^C$ is valid for $|r-r'| > 4$. At last, starting with (S120) as our initial condition, $||\overline{Q_0(t)}\rangle\rangle$ is given by an analogue of the recursive expression (S80), with $\langle\langle\Theta\bar{\Theta}_{r,r'}|| \to \langle\langle\Theta\tilde{\Theta}_{r,r'}||$ and $M_{r,r'} \to M_{r,r'}^C$, alongside (S124). Although the details surrounding both C-ST and NC-ST are different, the overall structure appearing in the dynamics of $||\overline{Q_0(t)}\rangle\rangle$ is similar. Namely, the dependence on the nearest and second nearest neighbours is the same almost everywhere, i.e. $M_{r,r'}^C = M_{r,r'}$ for $|r-r'| > 4$.

## S.2. Prescription to obtain $\overline{C_2(r,t)}$

Now that we know the dynamics of $||\overline{Q_0(t)}\rangle\rangle$ we can obtain that of the OTOC.

For $t = 0$ (S82) holds. For $t \geq 1$ we can decompose $\overline{C_2(r,t)}$ using (S124) to obtain

$$\overline{C_2(r,t)} = \langle\langle Q_r || S ||\overline{Q_0(t)}\rangle\rangle = \langle\langle Q_r || S \left( \sum_{R,R'=0}^{L/2-1} \frac{1}{8(2+\delta_{r,r'})} ||\Theta\tilde{\Theta}_{2R,2R'}\rangle\rangle \langle\langle\Theta\tilde{\Theta}_{2R,2R'}||\overline{Q_0(t)}\rangle\rangle \right)$$

$$= \frac{1}{6} \sum_{R=0}^{L/2-1} (\delta_{r,2R} + \delta_{r,2R+1}) \langle\langle\Theta\tilde{\Theta}_{2R,2R}||\overline{O_0^2(t)}\rangle\rangle. \tag{S127}$$

Note that, as we already suggested, the term $||\tilde{\theta}_{r,r'}\rangle\rangle \otimes ||\tilde{\pi}\rangle\rangle$ coming from (S124) does not contribute to the OTOC, since the particle-hole sector of $S||Q_r\rangle\rangle$ is $||\Pi + \pi\rangle\rangle$ and $\langle\langle\Pi + \pi||\tilde{\pi}\rangle\rangle = 0$.

Defining

$$K_{r,r'}^C(t) = \frac{1}{6} \langle\langle\Theta\tilde{\Theta}_{r,r'}||\overline{Q_0(t)}\rangle\rangle, \tag{S128}$$

analogous to (S84), the OTOC is then $\overline{C_2(2R,t)} = \overline{C_2(2R+1,t)} = K_{2R,2R}^C(t)$. Finally, we can use (S120) to get the OTOC at $t = 1$,

$$\overline{C_2(2R,t=1)} = \overline{C_2(2R+1,t=1)} = \frac{2}{9}\left(\delta_{R,-1} + \delta_{R,0}\right), \tag{S129}$$

and, at subsequent time steps, the dynamics of $\overline{C_2(r,t)}$ is determined by that of $\langle\langle\Theta\tilde{\Theta}_{r,r'}||\overline{O_0^2(t)}\rangle\rangle \Leftrightarrow K_{r,r'}^C(t)$ we just saw to be given by (S80) with the changes $\langle\langle\Theta\bar{\Theta}|| \to \langle\langle\Theta\tilde{\Theta}||$ and $M \to M^C$. This is, for $t > 1$ the OTOC is determined by (S86) with the changes $K_{r,r'}(t) \to K_{r,r'}^C(t)$ and $M_{r,r'} \to M_{r,r'}^C$, given by (S118) and (S126), respectively.

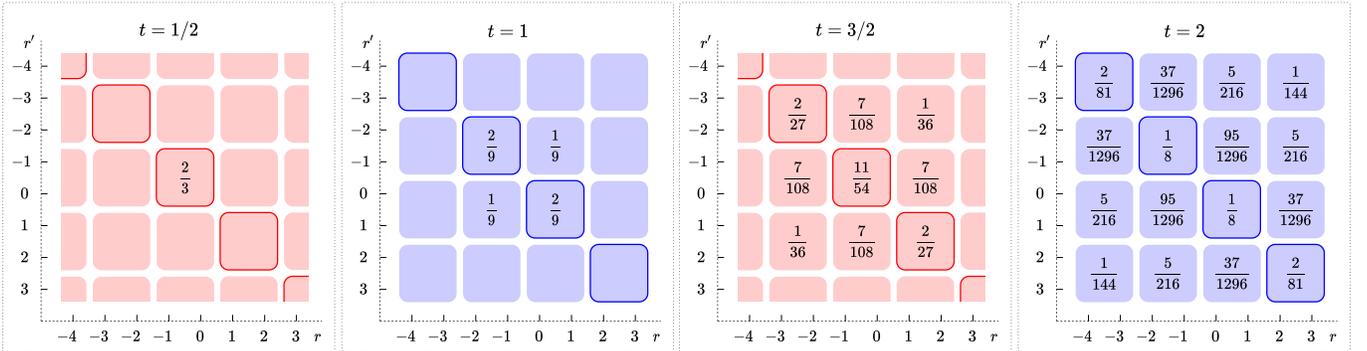

FIG. S5: The values in the boxes correspond to $K_{r,r'}^C(t)$ in the $r-r'$ plane, shown here for $t \in \{1/2, 1, 3/2, 2\}$ (empty boxes have zero value). Starting with $K_{r,r'}^C(t) = 2\delta_{r,r',0}$ as the initial condition, we evolve this by applying (S78) successively (with $\langle\langle\Theta\bar{\Theta}|| \to \langle\langle\Theta\tilde{\Theta}||$ and $M \to M^C$, given by (S118) and (S126), respectively). The highlighted diagonal gives $\overline{C_2(r,t)}$ for the C-ST case. This picture is equivalent to Fig. S4, valid for NC-ST and NC-T.

## S.3. Continuum limit of $\overline{C_2(r,t)}$

Since the bulk evolution of $K_{r,r'}^C(t)$ is equal to that of $K_{r,r'}(t)$, i.e. $m_{r,r'}^C|_{|r-r'|>2} = m_{r,r'}|_{|r-r'|>2}$, the continuum limit of the OTOC for the C-ST case is analogous to the one obtained for NC-ST. The differences of the diagonal with regards to the NC-ST case translate to

$$c \to c^C = 3/2, \tag{S130}$$

$$m'_{r,r} \to m'^C_{r,r} = \frac{1}{6}\begin{pmatrix} 2 & 1 \\ 1 & 2 \end{pmatrix}, \tag{S131}$$

$$f \to f^C = 2. \tag{S132}$$

In this case, the anisotropy of $m'^C_{r,r}$ leads to diffusion within the diagonal with weighting factor $f^C = 2$.

This results in the OTOC being given by (S91), with the changes $c = \sqrt{3}/2 \to c^C = 3/2$ and $f = 1/2 \to f^C = 2$. Thus, in the continuum limit, the OTOC is a Gaussian with broadening width $\sigma(t) = \sqrt{t}$, equal to the NC-ST and NC-T cases apart from the normalization which is $A(t) \to A^C(t) = f^C/f A(t) = f^C A/(2\sqrt{2\pi t})$.

## XIII. CONTINUUM LIMIT OF $\overline{\mathcal{C}(r,t)}$

Finally, we can take into account the contributions of $\overline{C_1(r,t)}$ and $\overline{C_2(r,t)}$ in the continuum limit, given respectively by (S54) and (S91), to obtain

$$\overline{\mathcal{C}(2R,t)} = \overline{\mathcal{C}(2R+1,t)} = \frac{1}{2^4}\left[\frac{A}{\sqrt{4\pi t}}\exp\left(-\frac{(2R)^2}{4t}\right) - f\frac{A}{4\pi t}\exp\left(-\frac{(2R)^2}{2t}\right)\right], \quad (S133)$$

where $A = 2$ and $f = 1/2$ for the NC-ST and NC-T cases and $f \to f^C = 2$ for the C-ST case.

Although $\overline{\mathcal{C}(r,t)}$ is typically given by the sum of two Gaussians with different standard deviation, in the infinite time limit it reduces to the first one, i.e. $\lim_{t\to\infty}\overline{\mathcal{C}(r,t)} = \overline{C_1(r,t)}/2^3$ since $\lim_{t\to\infty}\overline{C_2(r,t)}/\overline{C_1(r,t)} = 0$. This occurs because the normalization of $\overline{C_1(r,t)}$ is constant while that of $\overline{C_2(r,t)}$ is $\propto 1/\sqrt{t}$.

## XIV. NUMERICS

In the previous sections, we obtained analytical expressions for the TOC and OTOC for three instances of free fermion evolution: NC-ST, C-ST and NC-T. In the following, we compare these results with the respective approximate expressions in the continuum limit and also with data obtained from simulations. We also realize a fourth instance of free fermion evolution numerically − a temporally homogeneous case where randomness appears only along the space direction (NC-S). While the analytical results take into account all possible evolutions of the system, i.e. all possible circuits, the simulations approximate this by an average over $N_r = 4000$ disorder realizations. We look at a system with $L = 100$ using $C_1(r,0) = C_2(r,0) = 2\delta_{r,0}$ as the initial condition.

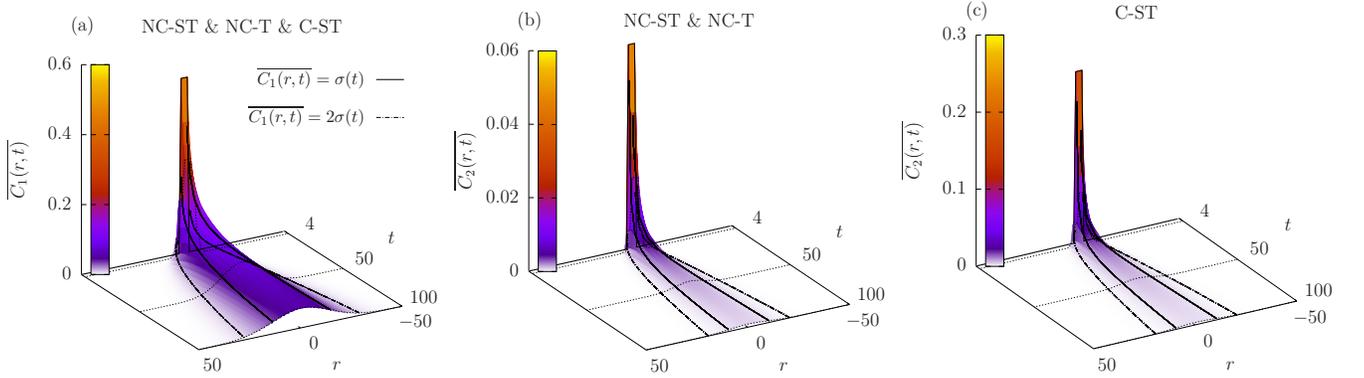

FIG. S6: (a) Height and colour maps of the exact $\overline{C_1(r,t)}$ in the $r-t$ plane for the NC-ST, NC-T and C-ST cases, given by (S52). (b) Exact $\overline{C_2(r,t)}$ for the NC-ST and NC-T cases, given by (S85) for $t = 1$ and (S86) for $t > 1$. (c) Exact $\overline{C_2(r,t)}$ for the C-ST case, given by (S129) for $t = 1$ and by an expression equivalent to (S86) for $t > 1$, with the changes $K \to K^C$ and $M \to M^C$ given respectively by (S128) and (S126). These results are for a system with $L = 100$ and periodic boundary conditions. The results are shown for $1 \leq t \leq 100$. The black curves indicate $\sigma(t)$ and $2\sigma(t)$, with $\sigma(t) = \sqrt{2t}$ for $\overline{C_1}$ and $\sigma(t) = \sqrt{t}$ for $\overline{C_2}$.

We begin with the analytical results, plotted in Fig. S6. We see that the profile of $\overline{C_1(r,t)}$ and $\overline{C_2(r,t)}$ at fixed time steps is given by a Gaussian whose width broadens with time according to $\sigma(t) = \sqrt{2t}$ and $\sigma(t) = \sqrt{t}$, respectively. We see that $\overline{C_1}$ dominates over $\overline{C_2}$. To compare these results with the data obtained from simulations we draw the profile of both $\overline{C_1(r,t)}$ and $\overline{C_2(r,t)}$ at fixed time steps.

Fig. S7 is a comparison of the analytic results and numerical simulation data for the three different cases NC-ST, NC-T and C-ST. The continuum limit in each case is tested by seeing whether $\overline{C_1}$ and $\overline{C_2}$ respect (S54) and

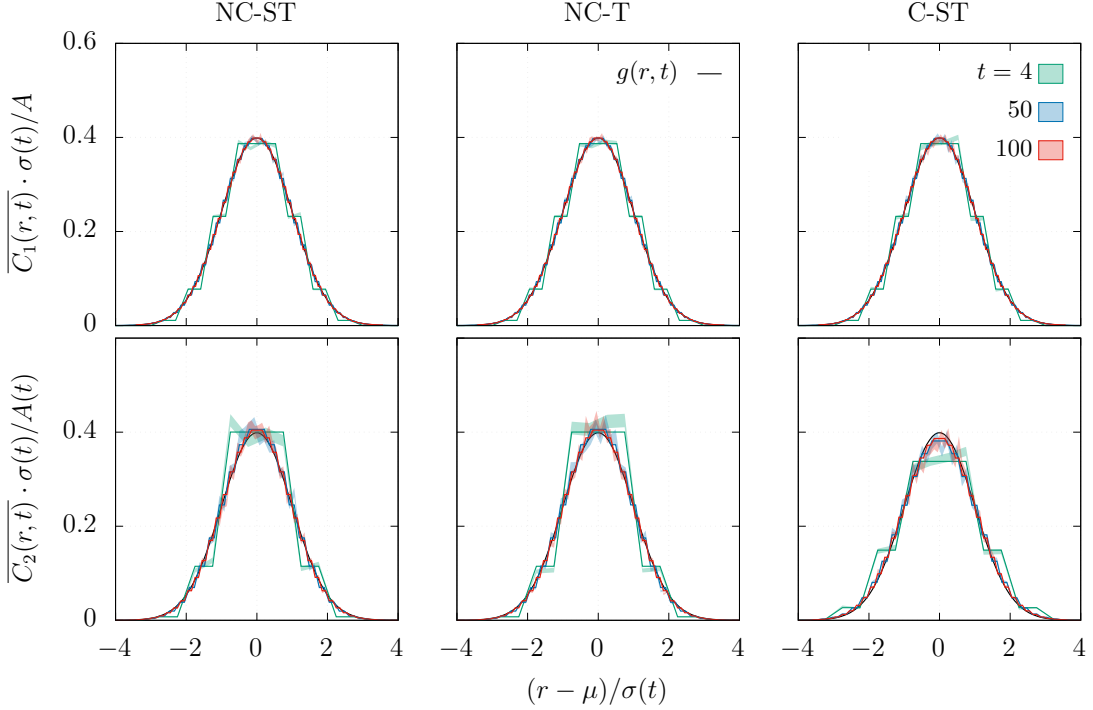

FIG. S7: Rescaled TOC, $\overline{C_1(r,t)} \cdot \sigma(t)/A$, (top) and OTOC, $\overline{C_2(r,t)} \cdot \sigma(t)/A(t)$, (bottom) as a function of $(r-\mu)/\sigma(t)$ at three fixed time steps $t \in \{4, 50, 100\}$ for a system with $L = 100$ and for the three cases NC-ST (left), NC-T (centre) and C-ST (right). Full lines correspond to the exact calculations seen in Fig. S6 and the colored regions to data obtained from simulations averaging over $N_r = 4000$ disorder realizations. Also, $g(r,t) = \frac{1}{\sqrt{2\pi}} \exp\left(-\frac{(r-\mu)^2}{2\sigma^2(t)}\right)$. For each case, the appropriate standard deviation $\sigma(t)$ and normalization $A$ or $A(t)$ are used. For $\overline{C_1(r,t)}$, $A = 2$ and $\sigma(t) = \sqrt{2t}$ for all cases. For $\overline{C_2(r,t)}$, $\sigma(t) = \sqrt{t}$ for all cases, $A(t) = 1/(2\sqrt{2\pi t})$ for NC-ST and NC-T and $A(t) = 2/(\sqrt{2\pi t})$ for C-ST. The deviation $\mu = 1/2$ appears because the brickwall structure of the circuit leads the TOC and OTOC to be centred around $\mu$ instead of 0.

(S91), respectively. This translates to $\overline{C_1(r,t)} \cdot \sigma(t)/A$ and $\overline{C_2(r,t)} \cdot \sigma(t)/A(t)$ collapsing to the Gaussian $g(r,t) = \frac{1}{\sqrt{2\pi}} \exp\left(-\frac{(r-\mu)^2}{2\sigma^2(t)}\right)$, with $A$ or $A(t)$ and $\sigma(t)$ given for each case in the caption of Fig. S7. We see in Fig. S7 that, for small time steps ($t = 4$), the discrete data does not follow a Gaussian, as expected, while for higher time steps it does. Indeed, for a reasonable system size ($L = 100$), the continuum limit quickly becomes a good approximation for either the TOC or the OTOC. For time steps significantly larger than $t = 100$, periodic boundary conditions cause the discrete data to diverge from the continuum limit expectation.

We have established that the analytical results, the data from simulations and the continuum limit expressions are in good mutual agreement. Before proceeding to analyse the numerical results obtained for the NC-S case, we provide some indication of the fluctuations present in both $C_1(r,t)$ and $C_2(r,t)$ by showing a single realization for each case of FF evolution in Fig. S8.

### S.1. NC-S: randomness in space alone

We now present numerical results for NC-S for $L = 100$ and $N_r = 4000$. The circuit for NC-S is built as shown in Fig. S1 (c): we construct two random layers as usual (one even and one odd) and then apply them repeatedly, such that there is discrete time translation symmetry.

We see in Fig. S9 that for the NC-S case both the TOC and OTOC are localized in space, decaying exponentially around $r = 0$ (see panels (b) and (d)). This phenomenon, arising from the disordered landscape, is a discrete time analogue of Anderson localization. This picture is unlike the diffusive behaviour seen for quadratic fermion evolution and the usual ballistic behaviour. Furthermore, this hints that the diffusive behaviour observed for NC-ST and NC-T

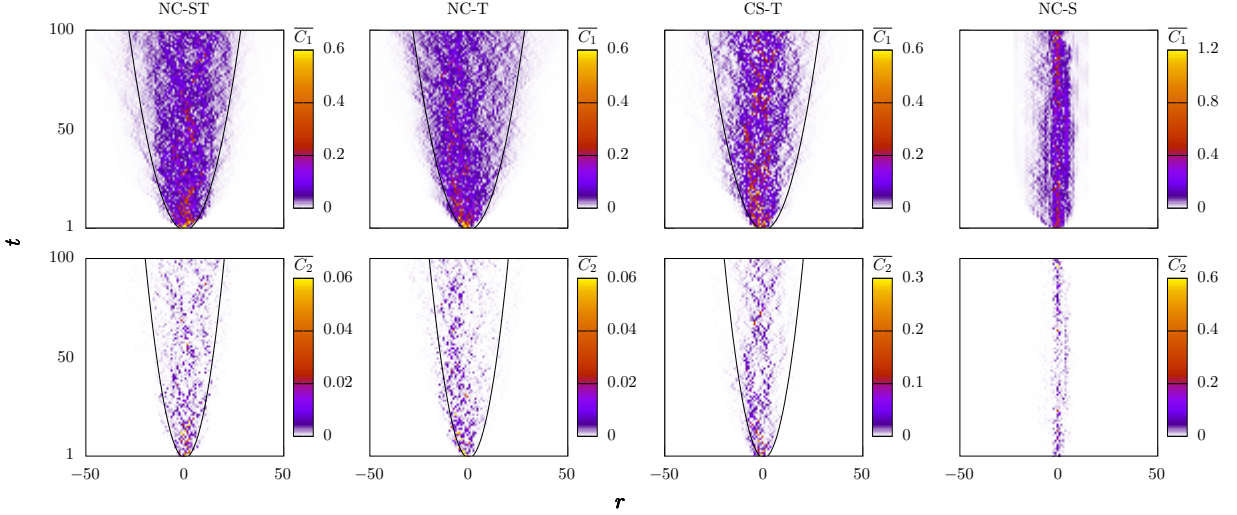

FIG. S8: Single disorder realization of $C_1(r,t)$ (top) and $C_2(r,t)$ (bottom) as a colour map in the $r-t$ plane for NC-ST (left), NC-T (centre left), CS-T (centre right) and NC-S (right). The black curves are $r - \mu = 2\sigma(t)$, i.e. $t = (r-\mu)^2/8$ for $\overline{C_1}$ and $t = (r-\mu)^2/4$ for $\overline{C_2}$. These results are for a system with $L = 100$ and periodic boundary conditions. The deviation $\mu$ appears as it did in Fig. S7.

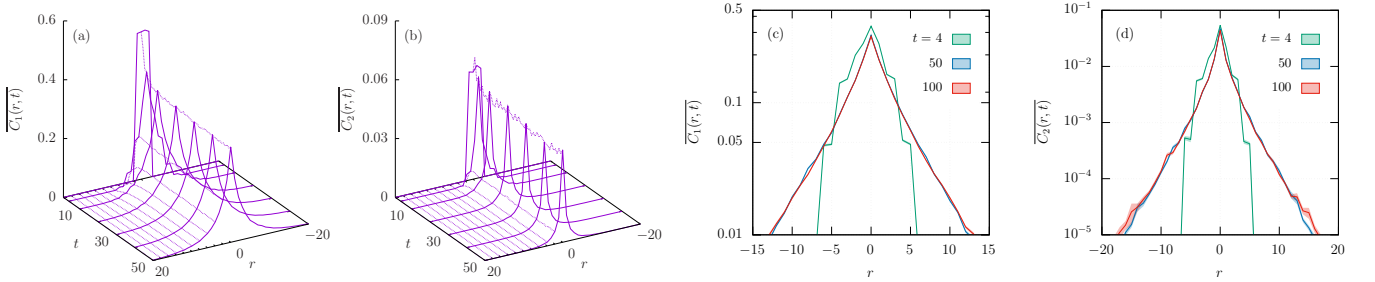

FIG. S9: (a) and (b) Height map of $\overline{C_1(r,t)}$ and $\overline{C_2(r,t)}$, respectively, in the $r-t$ plane for $1 \le t \le 50$, in the NC-S case. (c) and (d) Profile of $\overline{C_1(r,t)}$ and $\overline{C_2(r,t)}$ on a logarithmic scale, obtained by fixing $t$ to 4, 50 and 100 in panels (a) and (b). The data results from an average over $N_r = 4000$ simulations, for a system with $L = 100$ and periodic boundary conditions.

(and also C-ST) is due to randomness in time alone.

## XV. CONCLUSIONS

We have obtained analytical results showing that the TOC and OTOC, $\overline{C_1(r,t)}$ and $\overline{C_2(r,t)}$, spread diffusively in the $r-t$ plane for noisy free fermion evolution, with or without particle conservation. In total, three distinct cases were analysed − a particle non-conserving process with randomness in space and time (NC-ST), its particle conserving analogue (C-ST) and a noisy non-conserving case with spatial homogeneity (NC-T) − which differ either by the structure of the circuit or by the unitaries composing it (Fig. S1).

For the NC-ST case, the $\overline{C_1(r,t)}$ dynamics is similar to a classical random walk in one spatial dimension: although a single realization of $C_1$ does not describe a classical random walk, the averaged $\overline{C_1}$ is equivalent to an average over all possible random walks respecting the geometry of the circuit. Specifically, starting from $\overline{C_1(r,t=0)} = 2\delta_{r,0}$, each weight $\overline{C_1(r,t)}$ is divided among its nearest neighbours $\overline{C_1(r-1,t+1/2)}$ and $\overline{C_1(r+1,t+1/2)}$ (Section VII). While the dynamics of $\overline{C_1}$ is a 1D diffusive process, $\overline{C_2}$ is given by the diagonal of a quantity, $K_{r,r'}(t)$, which diffuses in 2D (Section VIII). The bulk behaviour of $K_{r,r'}(t)$ can also be described by an average over all possible classical random

walks, but in 2D, with each weight $K_{r,r'}(t)$ splitting into its nearest neighbours $K_{r\pm 1,r'\pm 1}(t+1/2)$. The NC-T case differs from NC-ST by the way the circuit is built − instead of having spatial disorder, a gate is randomly chosen at each half time step and applied to all pairs of sites (Fig. S1). In principle, this could lead the dynamics of $\overline{C_1}$ and $\overline{C_2}$ to arise from respectively 2D and 4D processes (Sections IX and X), but the initial condition reduces them to the 1D and 2D processes already known for NC-ST. To obtain the particle conserving dynamics (C-ST), the free fermion gates used in the NC-ST case, built from orthogonal matrices, are replaced by gates built out of unitary matrices. Since the average of two orthogonal and two unitary gates needed to obtain $\overline{C_1}$ coincide, the TOC is equal to the one obtained for the NC-ST case (Section XI). The picture is different for $\overline{C_2}$ − it is the only correlator which differs from the NC-ST results out of the cases studied. This happens because the average of four gates needed to compute $\overline{C_2}$ is different for orthogonal and unitary matrices (Section XII). However, this does not change the fundamental diffusive nature of the process. It merely leads to a different normalization in the continuum limit.

In the continuum limit, we saw in Section VII S.3 that $\overline{C_1(r,t)}$ is given by a Gaussian normalized to $A = 2$ whose standard deviation broadens in time with $\sigma(t) = \sqrt{2t}$ (for all cases) and $\overline{C_2(r,t)}$ is also given by a Gaussian, but with standard deviation $\sigma(t) = \sqrt{t}$ and whose normalization decreases in time according to $A(t) = fA/(2\sqrt{2\pi t})$, with $f = 1/2$ for NC-ST and NC-T (Section VIII S.3) and $f = 2$ for C-ST (Section XII S.3). Thus, since $A(t) \propto 1/\sqrt{t}$, $\overline{C_1}$ is the leading term in the continuum limit, i.e. $\lim_{t\to\infty} \overline{C_2(r,t)}/\overline{C_1(r,t)} = 0$.

Finally, let us address the connection between the single and many-body correlators. We established a relation between the two representations in Section IV: the many-body TOC and OTOC, $\mathcal{C}_1(r,t)$ and $\mathcal{C}_2(r,t)$, are mixtures of the single-body TOC and OTOC, $C_1(r,t)$ and $C_2(r,t)$, and vice versa. For our particular observables, the many-body TOC is constant, $\mathcal{C}_1(r,t) = 1/2^4$, and the non-trivial diffusive behaviour is encoded in the many-body OTOC: $\mathcal{C}_2(r,t) = \frac{1}{2^4}[1 + 2C_2(r,t) - 2C_1(r,t)]$.

---



\end{document}